\documentclass[acmtog]{acmart}

\usepackage{color}
\usepackage{multirow}
\usepackage{makecell}
\usepackage{tabularx}
\usepackage{booktabs} 
\usepackage{hyperref}
\hypersetup{
    colorlinks = true,
    linkcolor=blue,
    filecolor=blue,      
    urlcolor=blue,
    citecolor=cyan,
}

\usepackage{wrapfig}
\usepackage{hyphenat}
\usepackage{soul,color}
\usepackage{amsopn}
\usepackage{amsmath}
 
\usepackage{amssymb}
\usepackage{tabularx}
\usepackage{subfigure}

\definecolor{Red}{cmyk}{0,1,1,0}
\definecolor{Green}{cmyk}{1,0,1,0}
\definecolor{Cyan}{cmyk}{1,0,0,0}
\definecolor{Purple}{cmyk}{0.45,0.86,0,0}
\definecolor{Rosolic}{cmyk}{0.00,1.00,0.50,0}
\definecolor{Blue}{cmyk}{1.00,1.00,0.00,0}
\definecolor{BlueViolet}{cmyk}{0.86,0.91,0,0.04}
\definecolor{NavyBlue}{cmyk}{0.94,0.54,0,0}

\newcommand{\maNote}[1]{}

\usepackage[ruled]{algorithm2e} 

\SetAlFnt{\small}
\SetAlCapFnt{\small}
\SetAlCapNameFnt{\small}
\SetAlCapHSkip{0pt}

\acmJournal{TOG}

\setcopyright{rightsretained}
\acmJournal{TOG}
\acmYear{2022}\acmVolume{41}\acmNumber{4}\acmArticle{164}\acmMonth{7} \acmDOI{10.1145/3528223.3530086}

\begin{document}


\title{Artemis: Articulated Neural Pets with Appearance and Motion Synthesis }

\author{Haimin Luo}
\affiliation{%
  \institution{ShanghaiTech University}
  \country{China}}
\email{luohm@shanghaitech.edu.cn}

\author{Teng Xu}
\affiliation{%
  \institution{ShanghaiTech University}
  \country{China}}
\email{xuteng@shanghaitech.edu.cn}

\author{Yuheng Jiang}
\affiliation{%
  \institution{ShanghaiTech University}
  \country{China}}
\email{jiangyh2@shanghaitech.edu.cn}

\author{Chenglin Zhou}
\affiliation{%
  \institution{ShanghaiTech University}
  \country{China}}
\email{zhouchl@shanghaitech.edu.cn}

\author{Qiwei Qiu}
\affiliation{%
  \institution{ShanghaiTech University and Deemos Technology Co., Ltd.}
  \country{China}}
\email{qiuqw@shanghaitech.edu.cn}

\author{Yingliang Zhang}
\affiliation{%
  \institution{DGene Digital Technology Co., Ltd.}
  \country{China}
}
\email{yingliang.zhang@dgene.com}

\author{Wei Yang}
\affiliation{
  \institution{Huazhong University of Science and Technology}
  \country{China}}
\email{weiyangcs@hust.edu.cn}

\author{Lan Xu}
\affiliation{%
  \institution{ShanghaiTech University}
  \country{China}}
\email{xulan1@shanghaitech.edu.cn}

\author{Jingyi Yu}
\authornote{corresponding author}
\affiliation{
  \institution{ShanghaiTech University}
  \country{China}
}
\email{yujingyi@shanghaitech.edu.cn}

\renewcommand\shortauthors{Luo, H. et al}

\begin{abstract}
We, humans, are entering into a virtual era and indeed want to bring animals to the virtual world as well for companion. 
Yet, computer-generated (CGI) furry animals are limited by tedious off-line rendering, let alone interactive motion control. 
%
%
In this paper, we present ARTEMIS, a novel neural modeling and rendering pipeline for generating ARTiculated neural pets with appEarance and Motion synthesIS.
Our ARTEMIS enables interactive motion control, real-time animation, and photo-realistic rendering of furry animals.
The core of our ARTEMIS is a neural-generated (NGI) animal engine, which adopts an efficient octree-based representation for animal animation and fur rendering. The animation then becomes equivalent to voxel-level deformation based on explicit skeletal warping. We further use a fast octree indexing and efficient volumetric rendering scheme to generate appearance and density features maps.
Finally, we propose a novel shading network to generate high-fidelity details of appearance and opacity under novel poses from appearance and density feature maps.
For the motion control module in ARTEMIS, we combine state-of-the-art animal motion capture approach with recent neural character control scheme.
We introduce an effective optimization scheme to reconstruct the skeletal motion of real animals captured by a multi-view RGB and Vicon camera array. 
We feed all the captured motion into a neural character control scheme to generate abstract control signals with motion styles.
We further integrate ARTEMIS into existing engines that support VR headsets, providing an unprecedented immersive experience where a user can intimately interact with a variety of virtual animals with vivid movements and photo-realistic appearance.
Extensive experiments and showcases demonstrate the effectiveness of our ARTEMIS system in achieving highly realistic rendering of NGI animals in real-time, providing daily immersive and interactive experiences with digital animals unseen before.   
We make available our ARTEMIS model and dynamic furry animal dataset at \url{https://haiminluo.github.io/publication/artemis/}.

\end{abstract}

%
%


%
%


\begin{teaserfigure}
  \centering
\includegraphics[width=1.0\linewidth]{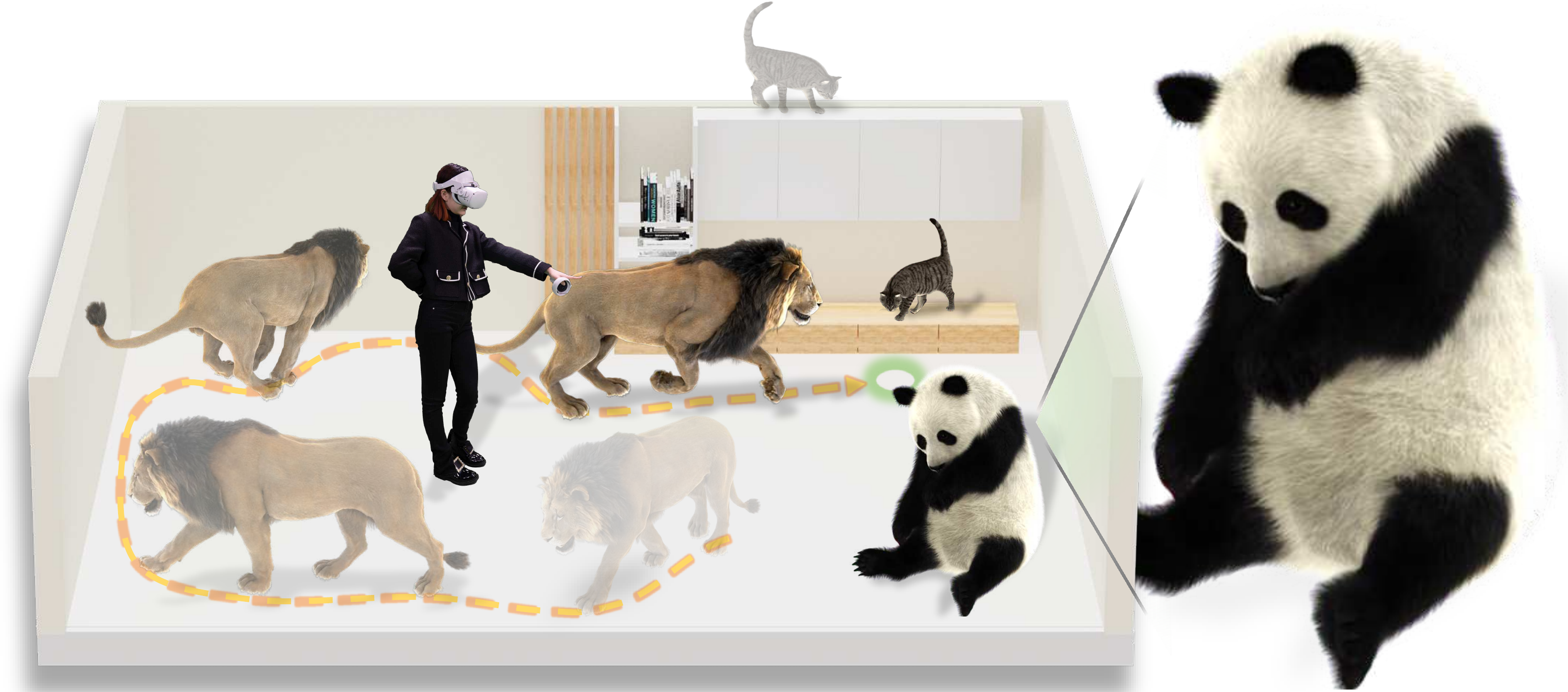}
  \caption{\textbf{Demonstration of our neural pet in a VR environment}. We present an approach, we call ``Artemis'',  for generating photo-realistic and interactable virtual pets. Traditional approaches animate animals with rigged mesh models and skeleton-based skinning techniques, but they can not handle furs well. Instead, we represent animals as an animatable neural volume and render animal appearance and furs in real-time (see the right figure for rendering results). Moreover, we use the motion synthesis approach, i.e., the Local Motion Phase, to generate skeletal motion for the animal according to the user's control signals. We demonstrate this concept in the left figure, where the user gives a destination point, and the virtual lion follows the skeleton and motion and moves to the destination.
  }
  \label{fig:teaser}
\end{teaserfigure}

\maketitle

\section{INTRODUCTION}
Our love for animals is a great demonstration of humanity. Digitization of fascinating animals such as Aslan the Lion in the Chronicles of Nardia to Richard Parker the Tiger in Life of Pie brings close encounters to viewers, unimaginable in real life. Despite considerable successes in feature films, accessing computer-generated imagery (CGI) digital animals and subsequently animating and rendering them at high realism have by far been a luxury of VFX studios where the process of creating them requires tremendous artists’ efforts and high computational powers, including the use of render farms. In fact, even with ample resources, photo-realistic rendering of animated digital animals still remains off-line and hence is not yet ready for primetime on Metaverse, where a user should be able to guide the movement of the animal, interact with it, and make close observations, all at high photo-realism in real-time. 

The challenges are multi-fold, but the conflict demand between real-time and photo-realistic rendering is at the core. First, animals are covered with furs traditionally modeled with hundreds of thousands of hair fiber/strands, using videos and photos as references. The traditional modeling process is tedious and requires exquisite artistic skills and excessive labor. The rendering process, on the other hand, is equally time-consuming: off-line ray tracing is typically adopted to produce photo-realism in translucency, light scattering, volumetric shadows, etc, and far from real-time even with the most advanced graphics hardware. 
In addition to rendering, animating the model at an interactive speed and with high realism remains challenging. For example, animations of animals in the latest remake of the Lion King were crafted by hand, based on the reference clips to match skeletal and muscle deformations. An automated tool is in urgent need for facilitating motion controls and even transfers (e.g., from dogs to wolves). Finally, for digital animals to thrive in the virtual world, they should respond to users’ instructions, and therefore, both the rendering and animation components need tight integration with interaction.

In this paper, we address these critical challenges by presenting \textit{ARTEMIS}, a novel neural modeling and rendering framework for generating \textit{ARTiculated neural pets with appEarance and Motion synthesIS}. In stark contrast with existing off-line animation and rendering systems, ARTEMIS supports interactive motion control, realistic animation, and high-quality rendering of furry animals, all in real-time. Further, we extend ARTEMIS to OpenVR to support consumer-level VR headsets, providing a surreal experience for users to intimately interact with various virtual animals as if in the real world (see Fig.~\ref{fig:teaser}).

\begin{figure*}
	\includegraphics[width=\linewidth]{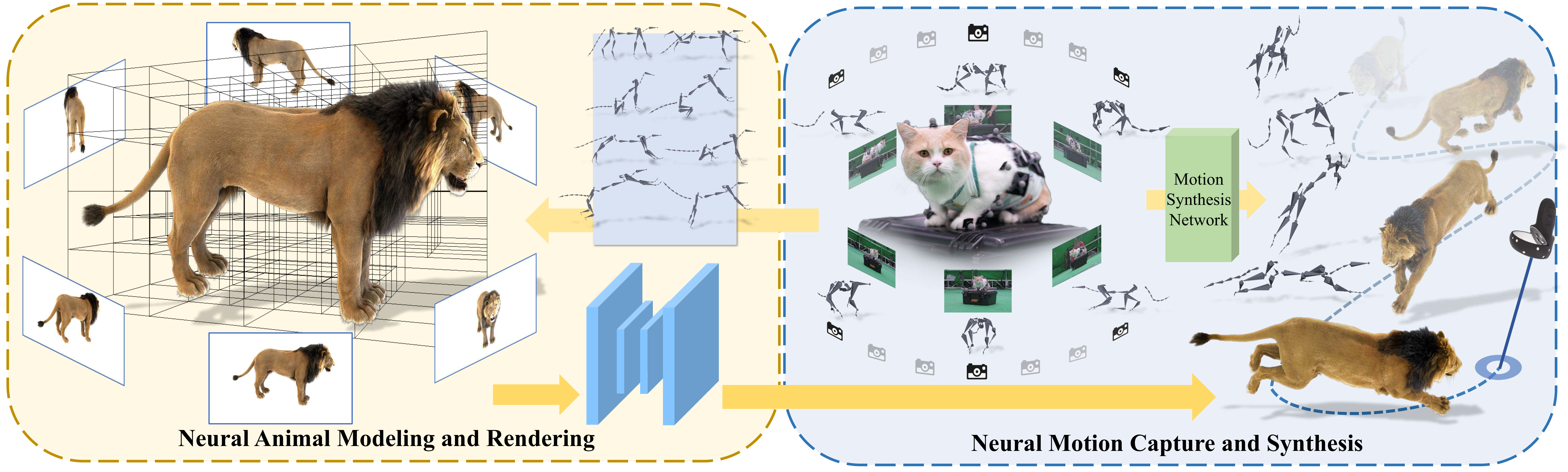}
	\caption{\textbf{Overview of our ARTEMIS system for generating articulated neural pets with appearance and motion synthesis.} ARTEMIS consists of two core components. In the first module, given skeletal rig and skinning weights of CGI animal assets and corresponding multi-view rendered RGBA images in representative poses, we build a dynamic octree-based neural representation to enable explicit skeletal animation and real-time rendering for dynamic animals, which supports real-time interactive applications;  In the second module, we build a hybrid animal motion capture system with multi-view RGB and VICON cameras to reconstruct realistic 3D skeletal poses, which supports to training a neural motion synthesis network to enable a user to interactively guide the movement of the neural animals. The ARTEMIS system is further integrated into existing consumer-level VR headset platforms for immersive VR experience of neural-generated animals.}
	\label{fig:pipeline}
\end{figure*}

On appearance modeling and rendering, ARTEMIS presents a neural-generated imagery (NGI) production pipeline to replace the traditional realistic but time-consuming off-line CGI process to model animated animals. NGI is inspired by the Neural Radiance Field (NeRF) ~\cite{mildenhall2020nerf}, and various forms of its extensions for producing photo-realistic rendering of real-world objects ~\cite{ST-Nerf,su2021nerf,peng2021animatable, tretschk2021non}. These approaches, however, cannot yet handle elastic motions while maintaining visual realism. In our case, we train on CGI animal assets (dense multi-view RGBA videos rendered using Blender) under a set of pre-defined motion sequences, where each animal model contains a skeleton with Linear Blend Skinning (LBS) weights. Using Spherical Harmonics factorization, we embed the animal's appearance in a high-dimensional latent space. Specifically, we follow the same strategy as PlenOctree~\cite{yu2021plenoctrees} and build a canonical voxel-based octree for the whole sequence while maintaining the latent descriptors and the density field. Under this design, we can apply LBS skeletal warping to locate features under the canonical pose on live poses. To ensure image quality, we further add a convolutional decoder to enhance spatial texture details of furs. The complete process can be trained as a differentiable renderer where the result enables the real-time, photo-realistic, free-viewpoint rendering of dynamic animals.

Besides rendering, ARTEMIS provides realistic motion synthesis so that a user can control the animal in virtual space. We combine state-of-the-art animal motion capture techniques with the recent neural character control ~\cite{starke2020local}. Since there still lacks a comprehensive 3D mocap database, we collect several new datasets using a hybrid RGB/Vicon mocap dome. Similar to human pose priors ~\cite{SMPLX2019}, we pre-train an animal pose prior from the data via Variational AutoEncoder (VAE) and use it as a regularizer for motion estimation. The resulting motion sequences are used to drive neural character animations~\cite{holden2017phase, zhang2018mann, starke2019neural,starke2020local,starke2021neural}. In particular, we employ the local motion phase (LMP) technique ~\cite{starke2020local} to guide the movement of the neural animal under common commands by real-world pet owners.
Since both neural rendering and motion controls of ARTEMIS achieve real-time performance, we further integrate ARTEMIS into existing consumer-level VR headset platforms. Specifically, we fine-tune our neural training process for binocular rendering and develop a hybrid rendering scheme so that the foreground NGI results can be fused with the rasterization-produced background. Comprehensive experiments show that ARTEMIS produces highly realistic virtual animals with convincing motions and appearance. Coupled with a VR headset, it provides a surreal experience where users can intimately interact with a variety of digital animals in close encounters.

To summarize, our main contributions include:
\begin{itemize} 
	\setlength\itemsep{0em}
	\item We propose a novel neural modeling and rendering system, ARTEMIS, that supports natural motion controls and user interactions with virtual animals on 2D screens or in VR settings. In particular, we collect a new motion capture dataset on animals of different scales and train a new skeleton detector.

	\item We present a differentiable neural representation tailored for modeling dynamic animals with furry appearances. Our new approach can effectively convert traditional CGI assets to NGI assets to achieve real-time and high-quality rendering. 
	
	\item We provide controllable motion synthesis schemes for various NGI animals under the VR setting. A user can send commands to or intimately interact with the virtual animals as if they were real.
	
\end{itemize}

\begin{figure*}[t]
  \includegraphics[width=\linewidth]{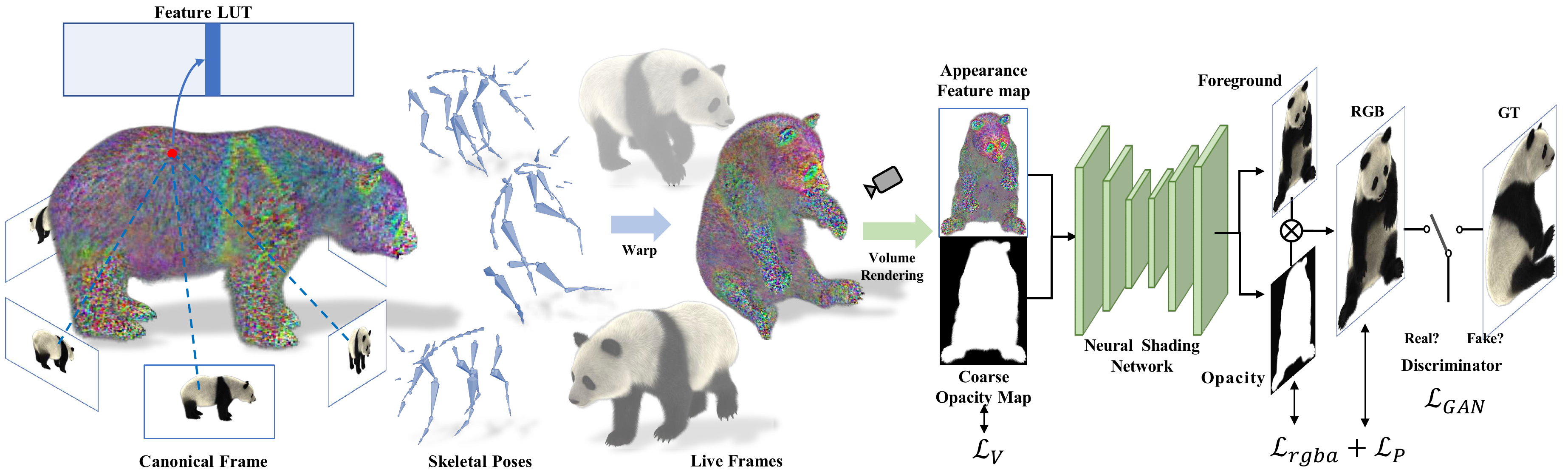} 
  \caption{\textbf{The algorithm pipeline of our Animatable Neural Animals.} Given multi-view RGBA images of traditional modeled animals rendered in canonical space, we first extract a sparse voxel grid and allocate a corresponding feature look-up table as a compact representation, together with an octree for quick feature indexing. Then we pose the character to training poses using the rig of the traditional animal asset and conduct efficient volumetric rendering to generate view-dependent appearance feature maps and coarse opacity maps. We next decode them into high-quality appearance and opacity images with the convolutional neural shading network. We further adopt an adversarial training scheme for high-frequency details synthesis.}
  \label{fig:neural_rendering_pipeline}
\end{figure*}

\section{RELATED WORK}

\subsection{Fuzzy object modeling and rendering}
Traditional methods focus on physical simulation-based and image-based modeling to model fuzzy objects like human hair and animal fur. 
Physical simulation-based modeling methods directly represent hair as existing physical models.
Hair rendering can date back to texel object method~\cite{KajiyaKay1989RenderingFur}, which proposed anisotropic lighting model object to be rendered as furry surfaces. This model is still used in some modern CG rendering engines. 
Light scattering model~\cite{Marschner2003LightScattering} of hair is then proposed, which modeling hair lighting with two specular layer effects, and is improved by considering energy-conserving~\cite{dEon2011EnergyConserving}.
A more complex microcosmic hair model is proposed, which adds a medulla kernel in hair and fur~\cite{Yan2015PhysicallyAccurateFur} and is accelerated by directly performing path tracing~\cite{Chiang2015PathTracingFur}.
Mass spring model~\cite{Selle2008MassSpringModel} is proposed to simulate hair which represents straight hair better.
Floating Tangents algorithm~\cite{DJ2013FloatingTangents} approximates hair curves with helices model.
Grid hair model starts from NVIDIA HairWorks technology and is improved to mesh model which provides direct control of overall hair shape ~\cite{Yuksel2009HairMeshes}.
On the other hand, image-based modeling methods try to reconstruct hair models directly from multi-view or single view images.
Multi-view images based hair reconstruction relies on direction field and hair tracking growth algorithm~\cite{Paris2004MultiHairCap, Paris2008HairPhotobooth, Wei2005MultiHairModel, Luo2013SHC, Bao2016HairModelOrientation}.
Single view image-based hair reconstruction, which relies on direction field and needs hair guideline as input~\cite{Chai2015SingleHighHair, Ding2016SingleHairHelix}, the data-driven based method rely on hair model database to perform hair information matching~\cite{Hu2015SingleHairData, Chai2016AutoHair}.

Besides human hair, other methods focus on reconstructing furry objects.
IBOH uses the multi-background matting method to construct opacity hull, which represents furry objects as opacity point clouds~\cite{Matusik2002IBOH}.
Neural representation methods are also proved to perform well on furry objects like hair and fur. NOPC method implements a neural network algorithm to get similar opacity point clouds in neural network representation~\cite{Wang2020NOPC}, which can efficiently generate high-quality alpha maps even with low-quality reconstruction from small data. The latter work ConvNeRF changes point clouds representation to opacity radiance field~\cite{9466273}, which enables high quality, global-consistent, and free-viewpoint opacity rendering for fuzzy objects.

Recall all furry objects modeling method, it is seen that traditional physical modeling and rendering method is able to generate high-quality method, but cost much times computation than simple mesh due to the complex ray reflection property of hair and fur, which is difficult to perform real-time rendering. Image-based methods directly reconstruct hair model from images, which is more efficient to generate static hair model but lose view consistency and cost much on dynamic scene effects. Recent neural representation methods perform well on fuzzy object modeling and rendering but still cannot generate real-time dynamic hair and fur effects.

\subsection{Virtual animals}

Animal parametric models have gradually been investigated in recent years, corresponding to general human parametric models. A series of works are proposed to capture and reconstruct animals like dogs ~\cite{biggs2020left}, horses~\cite{zuffi2019three}, birds ~\cite{kanazawa2018learning} and other medium animals ~\cite{cashman2012shape,zuffi20173d, biggs2018creatures}. The seminal work ~\cite{cashman2012shape} of Cashman and Fitzgibbon learns a low-dimensional 3D morphable animal model by estimating the shape of dolphins from images. This approach was limited to specific classes(dolphins, pigeons) and suffered from an overly smooth shape representation. ~\cite{kanazawa2016learning} learns a model to deform the template model to match hand clicked correspondences. However, their model is also animal-specific. SMAL ~\cite{zuffi20173d} extends SMPL~\cite{loper2015smpl} model to the animal domain; they create a realistic 3D model of animals and fit this model to 2D data, overcoming the lack of motion and shape capture for animal subjects with prior knowledge. ~\cite{kanazawa2018learning} combine the benefits of the classically used deformable mesh representations with a learning-based prediction mechanism. ~\cite{Kearney_2020_CVPR} built the first open-source 3D motion capture dataset for dogs using RGBD cameras and Vicon optical motion capture system. Comparably, our method builds a hybrid capture system with marker-based capture for docile and small pets(e.g., dogs and cats) and only RGB footage for dangerous and large animals. Besides, we reconstruct the realistic 3D skeletal poses for each kind of animal above.

\subsection{Neural Modeling and Rendering}

Various explicit representations have been incorporated into the deep learning pipeline for geometry and appearance learning via differentiable renderings, such as pointclouds~\cite{insafutdinov18pointclouds, lin2018learning, roveri2018network, yifan2019differentiable, aliev2020neural, wu2020multi, kolos2020transpr, FewshotNHR_IJCAI2021}, textured meshes~\cite{chen2019learning, kato2018neural, liu2020NeuralHumanRendering, liu2019neural, shysheya2019textured, 10.1145/3450626.3459749} and volumes~\cite{mescheder2019occupancy, peng2020convolutional, 10.1145/3306346.3323020, sitzmann2019deepvoxels}. However, they suffer from holes, topological changes, or cubic memory footprint. 
Recent implicit representations employ Multi-Layer Perceptrons (MLP) to learn a continuous implicit function that maps spacial locations to distance field~\cite{park2019deepsdf, chabra2020deep, jiang2020local}, occupancy field~\cite{mescheder2019occupancy, peng2020convolutional, saito2019pifu, huang2020arch, he2021arch++} or radiance field ~\cite{mildenhall2020nerf, martin2021nerf, liu2020neural, bi2020neural}. They can naturally handle complicated scenes, but they usually suffer from high query time for rendering. 
For real-time rendering, explicit primitives~\cite{Lombardi21} or spheres~\cite{lassner2021pulsar} representations are employed, together with acceleration data structures~\cite{yu2021plenoctrees}. However, they require good initialization or high memory for dynamic scenes. MonoPort~\cite{li2020monocular, 10.1145/3407662.3407756} achieves impressive real-time reconstruction and rendering with an RGBD camera by combining PiFu~\cite{saito2019pifu} and octree for fast surface localization, but it does not support animation and suffers from insufficient rendering quality.

Recently NeRF~\cite{mildenhall2020nerf} has achieved impressive progress by learning radiance fields from 2D images. The following variants of NeRF aim to learn generalizable radiance fields~\cite{wang2021ibrnet, chen2021mvsnerf}, train with unposed cameras~\cite{wang2021nerf, yen2020inerf, meng2021gnerf}, model high-frequency details~\cite{tancik2020fourfeat, sitzmann2020implicit, 9466273} and opacity~\cite{9466273}, handle relighting and shading~\cite{bi2020neural, boss2021nerd, kuang2021neroic} or accelerate for real-time applications~\cite{neff2021donerf, garbin2021fastnerf, reiser2021kilonerf, hedman2021baking, yu2021plenoctrees}.
Most recent approaches extend NeRF to dynamic scenes by learning a deformation field~\cite{pumarola2021d, li2021neural, tretschk2021non, park2021nerfies, xian2021space} or training a hypernet~\cite{park2021hypernerf}. Such methods can only achieve playback or implicit interpolation. With estimated skeleton motion and parametric model, AnimatableNeRF~\cite{peng2021animatable} achieves pose control by inversely learning blending weights fields; however, it struggles for large motion due to the limited capacity of MLPs. NeuralBody~\cite{peng2021neural} proposes structured latent codes for dynamic human performance rendering and is then extended to be more pose-generalizable by adopting pixel-aligned features~\cite{kwon2021neural}. H-NeRF~\cite{xu2021h} constrains a radiance field by a structured implicit human body model to robustly fuse information from sparse views for better pose generalization. NeuralActor~\cite{liu2021neural} further introduces dynamic texture to neural radiance fields for better pose-dependent details. These works, however, still suffer from high query time consumption without acceleration.
Another line of research employs explicit volume~\cite{10.1145/3306346.3323020}, points~\cite{wu2020multi}, mesh~\cite{shysheya2019textured, 10.1145/3450626.3459749, liu2020NeuralHumanRendering, liu2019neural}, but suffers from low resolution representations. MVP~\cite{Lombardi21} achieves real-time fine detail rendering but still cannot realize pose extrapolation.

\section{Animatable Neural Animals}\label{sec:Renderer}
Animating digital animals have long relied on experienced artists. To produce photo-realistic appearance and opacity of furry animals, it is commonly a group effort to model tens of thousands of hair fibers and simulate their movements with physical soundness. Finally, real-time rendering is difficult to achieve even on high-end render farms. For interactive applications such as virtual pets, it is essential to simultaneously maintain photo-realism and real-time performance on both motion and appearance.

\paragraph{Neural Opacity Radiance Fields.}
In contrast to physically modeling translucency of furs, we adopt the neural rendering approach and cast the problem as view synthesis on the radiance field. The seminal work of the Neural Radiance Field conceptually provides a natural solution to fur rendering: NeRF represents a scene in terms of the color and density on each point along a ray where the density naturally reflects the opacity of the point. However, the alpha matte produced from NeRF tends to be noisy and less continuous than natural fur or hair. The recent ConvNeRF~\cite{9466273} addresses this issue by processing the image in feature space than directly in RGB colors. They formulate a neural opacity field as:%
\begin{equation}
    I = {\Phi} (F_p), \, \, F_p = \sum_i^{|\mathcal{S}|} T_i  \big ( 1 - \exp(-\sigma_i\delta_i) \big ) f_i
\end{equation}
where $\Phi$ is the opacity synthesis network, and $F_p$ is the aggregated image features, $T_i = \exp(-\sum_{j=1}^{i}\sigma_j\delta_j)$. $\mathcal{S}$ is the set of sampled points, and $\sigma_i$, $\delta_i$ are density and distance between sampled points, respectively. The key here is the use of feature $f_i$ to represent the opacity at each spatial point instead of directly using the density. The original ConvNeRF can only handle static objects. The brute-force approach is to conduct per-frame optimization for dynamic objects, which is clearly infeasible on neural animals that generally perform long motion sequences.

\begin{figure}[t]
  \includegraphics[width=\linewidth]{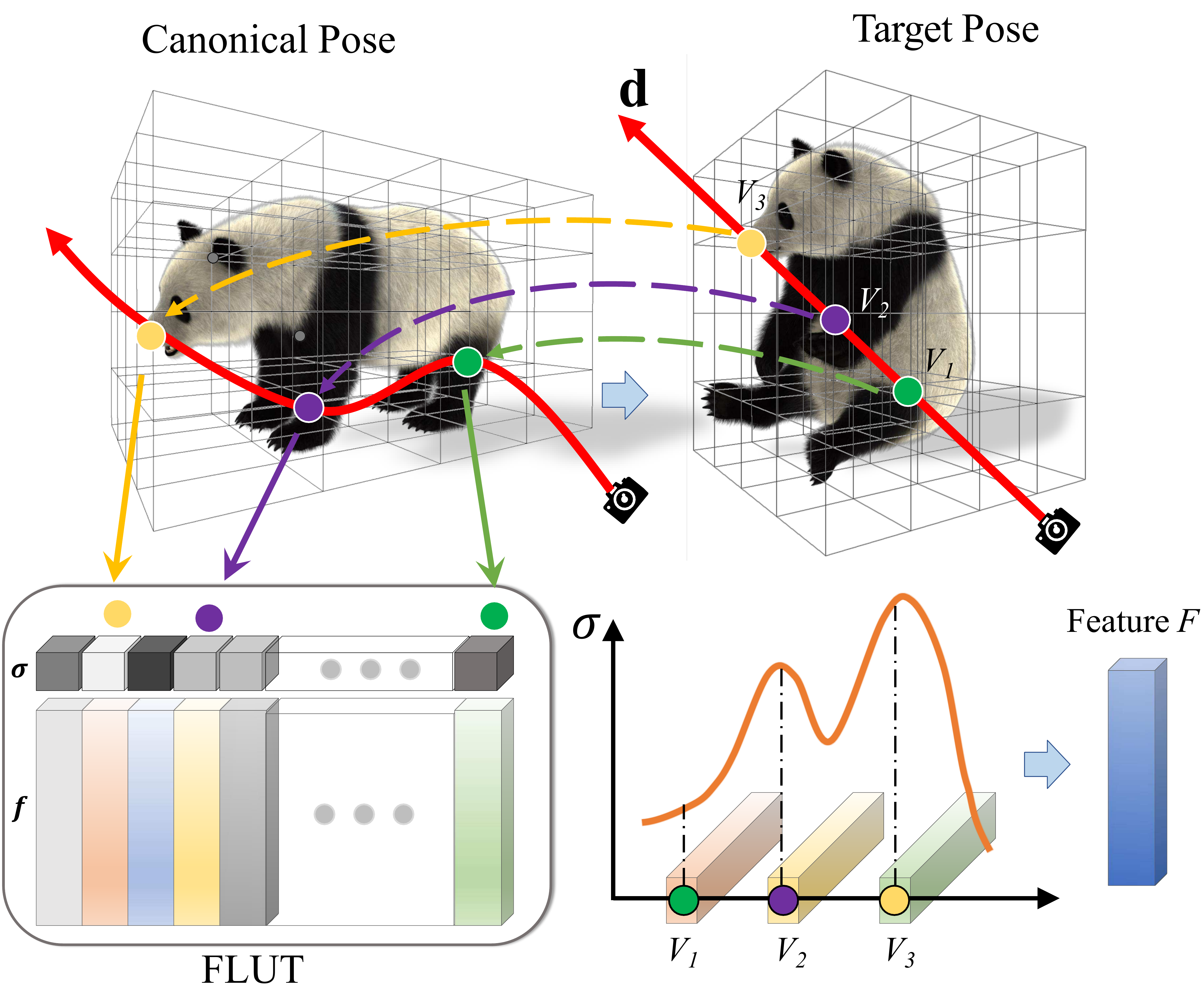}
  \caption{\textbf{Dynamic Animal Rendering Details}. Given a target skeletal pose, we deform the canonical index octree to live frames, then perform ray marching for efficient volume rendering to integrate the features along the ray to generate feature maps for the further rendering process.}
  \label{fig:render_details}
\end{figure}

\subsection{Animatable Neural Volumes} \label{Sec:ani_neural_volume}
We first extend the neural opacity radiance field to dynamic animals. Notice that our goal is not only to accelerate training but, more importantly, to realize real-time rendering. 
Inspired by PlenOctree~\cite{yu2021plenoctrees} for static scenes, we set out to store the opacity features in a volumetric octree structure for real-time rendering but under animations. Specifically, we devise a feature indexing scheme for quick feature fetching and rendering.
We further introduce a skeleton-based volume deformation scheme that employs skinning weights derived from the original CGI model to bridge the canonical frame with live frames for animation.
In addition, we devise a neural shading network to handle animated object modeling and adopt an efficient adversarial training scheme for model optimization.

\paragraph{Octree Feature Indexing.}
Unlike the original NeRF or PlenOctree, where the object's geometry is unknown, we have the CGI animal models as inputs. Therefore, we first convert a CGI animal character, such as a tiger or a lion, to the octree-based representation. Again, the original CGI model contains very detailed furs, and direct conversion onto discrete voxels can cause strong aliasing and significant errors in subsequent neural modeling. If we remove the furs and use only the bare model, the voxel representation will deviate significantly from the actual ones. Instead, we prefer "dilated" voxel representations that encapsulate the model. We apply a simple trick to resolve this issue: we use the rendered alpha matte from a dense set of views as inputs and conduct conservative volume carving to construct the octree: we initialize a uniform volume and carve it using the masks dilated from the alpha mattes. Notice, though, that we also need the rendered multi-view alpha matte later for training the neural opacity field. The resulting octree contains a voxel array $\mathbf{P}$ occupied in 3D space.
Using this volumetric representation, we aim to store a view-dependent feature $f$ at each voxel. We, therefore, allocate a separate data array $\textbf{F}$ called Features Look-up Table (FLUT) to store features and density values as in the original PlenOctree.
Here FLUT is used to efficiently query features at an arbitrary 3D location to accelerate training and inference.
For a given query point in space at the volume rendering process, we can index into FLUT in constant time and assign the corresponding feature and density to the point.

Inspired by PlenOctree~\cite{yu2021plenoctrees} that factorizes view-dep-\\endent appearance with Spherical Harmonics (SH) basis, we model our opacity feature $f$ also as a set of SH coefficients, i.e., $f = \{k_h^i\}_{h=1}^H$, where $k_h^i \in \mathbb{R}^C$ correspond to the coefficients for $C$ components, and $H$ is the number of SH functions. Given a query ray direction $d = (\theta, \phi)$, the view dependent feature $\textbf{S} \in \mathbb{R}^C$ can be written as:

\begin{equation}
    \textbf{S}(f, d) = \sum_{h=1}^H k_h^i Y_h(\theta, \phi)
\label{eq:Fv}
\end{equation} 
where $Y_h: \mathbb{S}^2 \to \mathbb{R}$ is the SH bases. In our implementation, we set the dimension of SH bases to 91.

\paragraph{Rigging and Deformation.} 
To make the octree adapt to animated animals, we directly "rig" the octree $\mathcal{V}_c$ in terms of its voxels from the canonical pose to the target skeleton $\mathcal{S}$, where $S$ is the skeleton parameter $\mathcal{S}=\{\textbf{r}, R, T\}$, $\textbf{r} \in \mathbb{R}^{J \times 3}$ is the joint rotation angles, $R \in \mathbb{R}^3$ is global rotation and $T \in \mathbb{R}^3$ is global translation.

To rig the octree under skeletal motion, the brute force way is to rig all voxels in terms of their distance to the skeletons. Following Linear Blending Skinning (LBS), we instead apply skinning weights to voxels using the skinned mesh provided by the CGI models.
Specifically, given mesh vertices and corresponding skinning weights, we generate per-voxel skinning weight by blending the weights of $m$ closest vertices $\{v_j\}_{j=1}^m$ instead of only one point~\cite{huang2020arch} on the mesh as:
\begin{equation}
    w(\textbf{p}_i) = \sum_{j=1}^{m}\alpha_j w_j, \,\, \alpha_j = e^{\delta_j} / \Sigma_k^m e^{\delta_k}
\end{equation}
where $w_j$ is the skinning weight of $v_j$ and $\delta_j = d_j - \mathop{min}_k^md_k$ with $d_j$ as the Euclidean distance between $v_j$ and voxel position $\textbf{p}_i$.

With the voxel set $\mathcal{V}_c$, their skinning weights $\mathcal{W}$ and a corresponding skeleton $\mathcal{S}_c$ ready, we then conduct deformation from the canonical pose $\mathcal{S}_c$ to the target pose $\mathcal{S}_t$ with the transformation matrices $M^c, M^t \in \mathbb{R} ^ {J\times4\times4}$ following LBS as: 
\begin{equation}
    \textbf{p}_i^t = \sum_{j=1}^{J}w_j(\textbf{v}_i) M^t_j (M^c_j)^{-1} \textbf{p}_i^c
\end{equation}
with canonical voxel position $\textbf{p}_i^c$ and $J$ joints.

\begin{figure}[t]
\centering
    \includegraphics[width=1.0\linewidth]{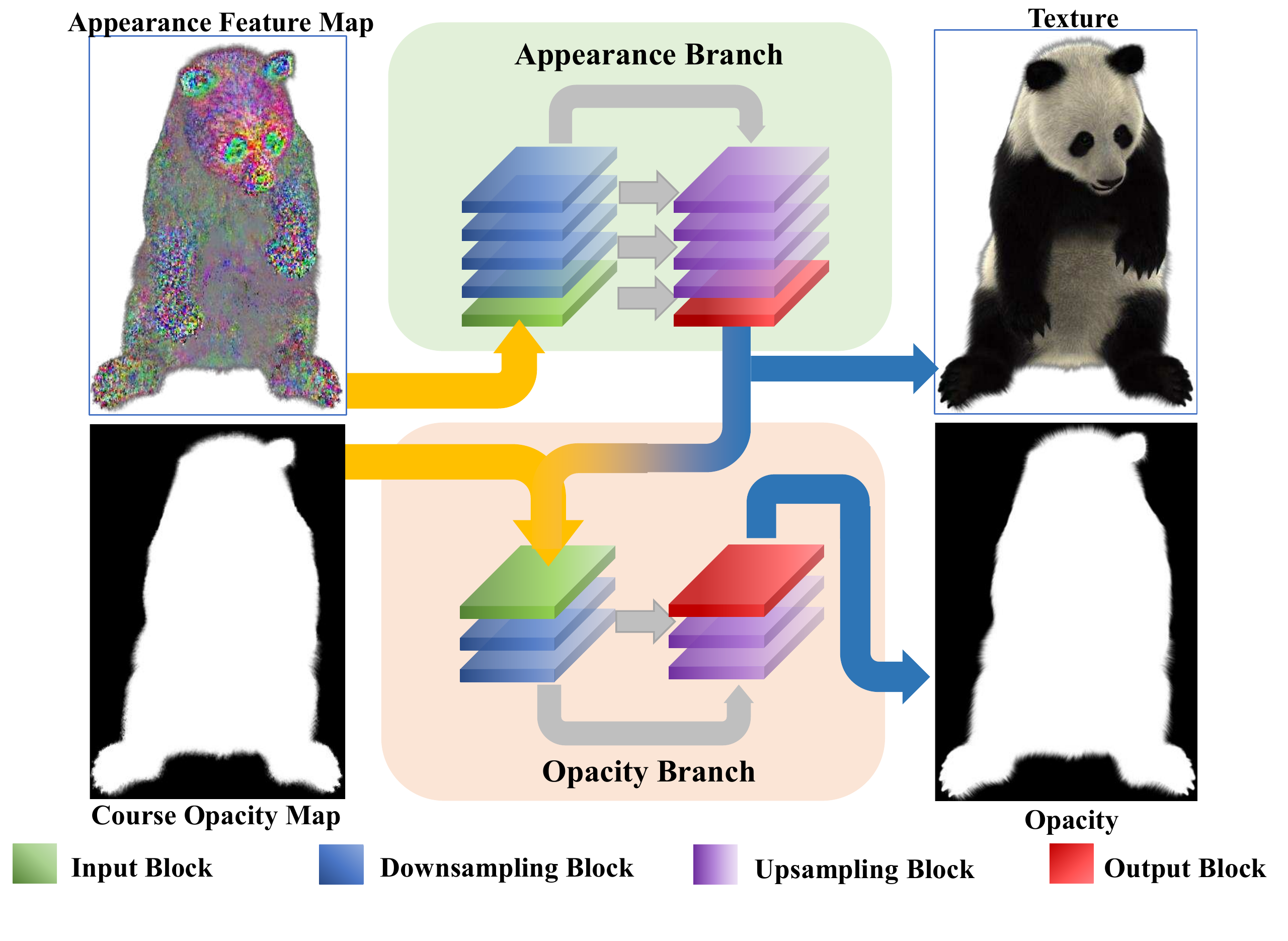}
\vspace{-7mm}
\caption{\textbf{Neural Shading Network.} Our four blocks appearance branch translates the feature map to the foreground texture image, while the two blocks opacity branch takes the texture image to refine the coarse opacity map.}
\label{fig:network}
\end{figure}

\paragraph{Dynamic Volume Rendering.}
Once we deform the octree to the target pose, we can render the neural animal via volume rendering and subsequently enable end-to-end training. We employ a differentiable volumetric integration scheme, as shown in Fig.~\ref{fig:render_details}. Given a ray $r_p^t = (\textbf{o}_p^t, \textbf{d}_p^t)$ with starting point $\textbf{o}_p^t$ and ray direction $\textbf{d}_p^t$ that corresponds to pixel $p$ viewing the volume under pose $\mathcal{S}_t$, we can compute the view-dependent feature of $p$ as:
\begin{equation}
    F_p^t = \sum_i^m\alpha_i \textbf{S}(f_i, \textbf{d}_p^c)
\end{equation} \label{eq:feature}
\begin{equation}
    \alpha_i = T_i \big (1 - \exp(-\sigma_i\delta_i) \big ), \,\,\,
    T_i =\exp(-\sum_{j=1}^{i}\sigma_j\delta_j)
\label{eq:alpha}
\end{equation}
where $\delta_i$ is the distance of the ray traveling through a voxel $\textbf{v}_i$, and $f_i$ is the corresponding SHs coefficients fetched using the index stored in $\textbf{v}_i$, $m$ is the number of hit voxels. 
To further preserve consistency over view-dependent effects, we map the ray directions to the canonical space with a rotation matrix $R_i^t$ by $\textbf{d}_p^c = (R_i^t)^{-1}\textbf{d}_p^t$.

Finally, we generate a view-dependent feature map $\mathcal{F}$ by applying Eqn.\ref{eq:feature} to each pixel and a coarse opacity map $\mathcal{A}$ by accumulating $\alpha_i$ along the rays. Since our neural animal model is dense and compact in space, uniform sampling used in NeuralVolumes \cite{10.1145/3306346.3323020} and NeRF \cite{mildenhall2020nerf} can lead to rendering empty voxels. To tackle this problem, we employ a highly optimized data-parallel octree construction technique~\cite{karras2012maximizing} that only uses around $10$ms  to build an octree for millions of voxels. With this speed, we manage to rebuild the octree under pose variations and perform a ray-voxel intersection test for dynamic scenes. 
We further apply an early-stop strategy based on the accumulated alpha, i.e., we stop the integration once alpha is greater than $1 - \lambda_{th}$ (we use 0.01 in our implementation). 

\paragraph{Neural Shading.}
We have by far rasterized the volume to a neural appearance feature map $\mathcal{F}$ and opacity $\mathcal{A}$ at the target pose and viewpoint. The final step is to convert the rasterized volume to a color image with a corresponding opacity map resembling a classical shader. 
To preserve high frequency details of the fur, it is essential to consider spatial contents in the final rendered image. Notice, though, that neither NeRF nor PlenOctree considers spatial correlation as all pixels are rendered independently. We instead adopt an additional U-Net architecture $\Phi$ following ConvNeRF~\cite{9466273} to perform image rendering. Note that our ray-marching-based sampling strategy enables full-image rendering, in contrast to the patch-based strategy of ConvNeRF. Precisely, our neural shading network $G$ consists of two encoder-decoder branches for RGB and alpha channels, respectively. 
The RGB branch converts $\mathcal{F}$ to texture images $\textbf{I}_f$ with rich fur details. The alpha branch refines the coarse opacity map $\mathcal{A}$ and $\textbf{I}_f$ to form a super-resolved opacity map $\mathbf{A}$. The process enforces multi-view consistency by explicitly utilizing the implicit geometry information encoded in $\mathcal{A}$.

\paragraph{Training.}
Recall that all features $f$ are stored in FLUT. For acceleration, we randomly initialize FLUT and then jointly optimize the dense feature array $\textbf{F}$ and the parameters of $G$ from the multi-view animal motion videos.
Our network aims to recover the appearances and opacity values of furry animals under free viewpoint. We therefore adopt the pixel-level $L_1$ loss as:
\begin{equation}
    \mathcal{L}_{rgba} = \frac{1}{N_P} \sum_{i}^{N_P} ( \|\hat{\mathcal{I}_i} - \mathcal{I}_i\|_1 + \|\hat{\alpha_i} - \alpha_i\|_1)
\end{equation}
where $\mathcal{I}, \alpha$ are ground truth color and alpha rendered from ground truth fuzzy CGI animal models, and $\hat{\mathcal{I}_i}, \hat{\alpha_i}$ are synthesized color and alpha values from our network. In particular, $\hat{\mathcal{I}_i}$ is the blended result using the predicted alpha matte. $N_P$ is the number of sampled pixels. 

To recover fine texture and geometry details, we employ the \textit{VGG19} perceptual loss~\cite{johnson2016perceptual}:
\begin{equation}
    \mathcal{L}_{P} = \frac{1}{N_I} \sum_{i}^{N_I} \sum_{l\in\{3,8\}} 
    (\|\phi^l(\hat{\textbf{I}}_i) - \phi^l(\textbf{I}_i)\|_1 + \|\phi^l(\hat{\textbf{A}}_i) - \phi^l(\textbf{A}_i)\|_1)
\end{equation}
where $\phi^l$ denotes the $l^{th}$ layer feature map of \textit{VGG19} backbone, and $N_I$ is the number of sampled images.

To encourage cross view consistency as well as maintain temporal coherency, we impose geometry priors encoded in the ground truth alpha maps on the coarse alpha map $\mathcal{A}$:
\begin{equation}
    \mathcal{L}_{V} = \frac{1}{N_I} \sum_{i}^{N_V} \|\mathcal{A}_i - \textbf{A}_i\|_1
\end{equation}

Under motion, voxels may be warped to same locations that occlude previous frames. This leads to voxel collisions and feature overwrites of a grid and breaks the temporal consistency. To address this issue, we propose a voxel regularization term (VRT)  that enforces features falling onto the same grid after warping should have identical values as:
\begin{equation}
    \mathcal{L}_{vrt} = \lambda_{vrt}\frac{1}{N_{v}}\sum_i^{N_v}\|\textbf{Q}(\textbf{p}_i^t, \textbf{T}^t) - \textbf{f}_i^c\|_1
\end{equation}
$\textbf{Q}$ is a query operator that outputs the corresponding feature of voxel at position $\textbf{p}_i^t$ in octree $\textbf{T}^t$. $N_v$ is the voxel number.

To further improve the visual quality of high-frequency appearance of furs, we employ the PatchGAN~\cite{isola2017image} discriminator $D$ to perform adversarial training with a GAN loss as:
\begin{equation}
    \mathcal{L}_{GAN} = \sum_i^{N_I} (\|D(\hat{\textbf{I}}_i)\|_2 + \|D(\textbf{I}_i) - 1\|_2)
\end{equation}

\begin{figure}[t]
  \includegraphics[width=\linewidth]{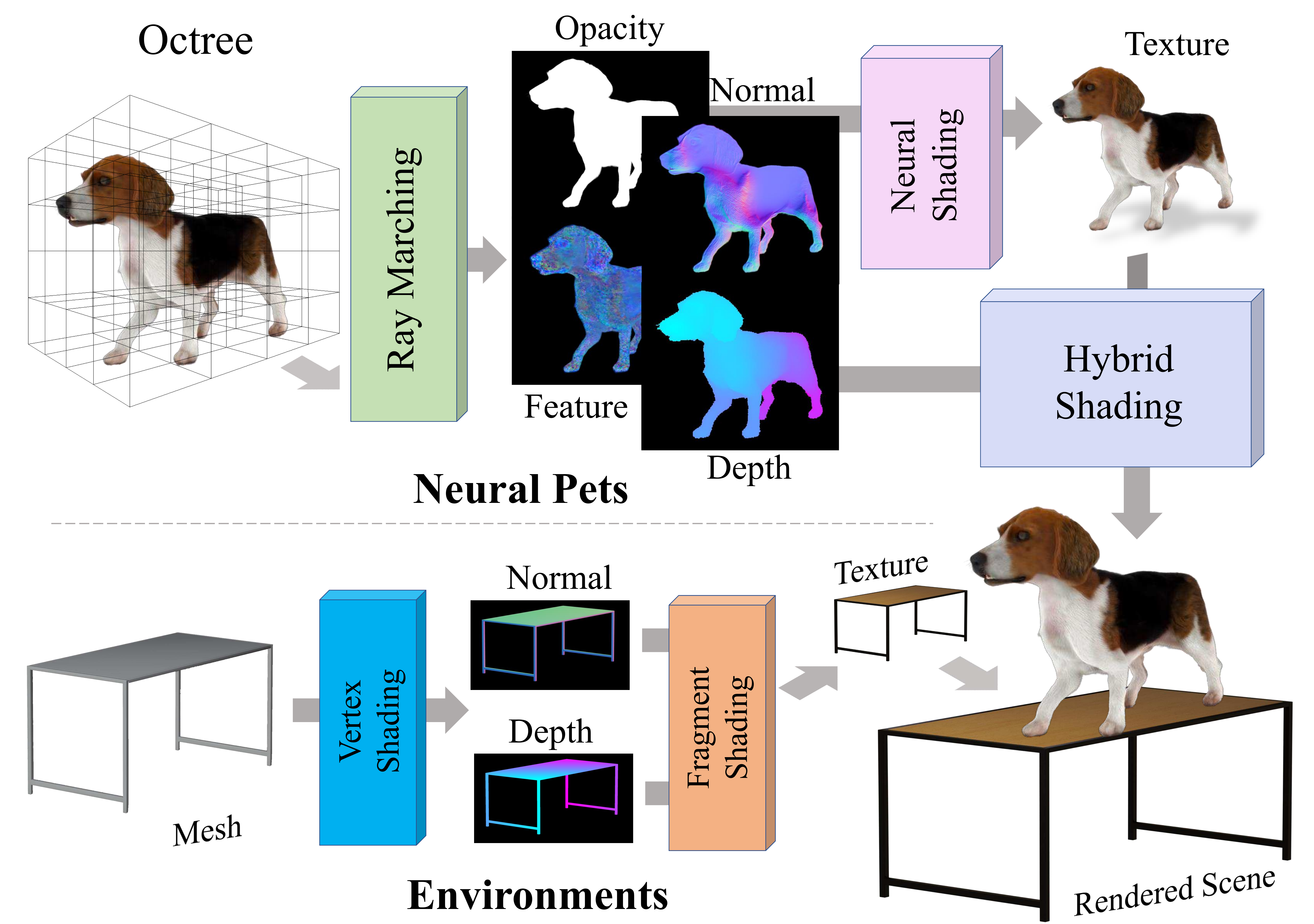}
  \caption{\textbf{Neural Rendering Engine.} We develop a neural rendering engine based on our animatable neural volumetric representation. It can render our neural animals into standard frame buffers (e.g., texture, alpha, depth) used in the traditional 3D rendering pipeline.}
  \label{fig:ray_marching}
\end{figure}

\paragraph{Hybrid Rendering.}
The processes of warping, octree construction, and volume rendering form a differentiable neural rendering pipeline and can be paralleled using C++ and CUDA C for efficient, end-to-end training. Once trained, the resulting octree structure supports real-time free-view rendering under animation. The solution can be further integrated with OpenGL and Unity engines, where a standard 3D rendering pipeline can conveniently render environments that host virtual animals.
It is essential to address shading and occlusions in order to correctly fuse the environment with the neural rendering results to provide a realistic viewing experience. We, therefore, extend the volume rendering process described in Sec.~\ref{Sec:ani_neural_volume} to render to different frame buffers. Firstly, the original outputs $\mathcal{F}$ and $\mathcal{A}$ correspond to neural textures for neural shading, and the final rendered images are stored in texture buffers. 
Neural rendered animals can also produce a depth buffer using a ray marching process: we cast a ray from each pixel from the image plane to our trained volumetric model, trace along the ray and locate the first voxel with a density larger than a threshold, and directly assign the distance as the depth. In contrast, we observe that directly applying such ray marching schemes on NeRF produces large errors near the boundary of the object, particularly problematic for furry objects. The rendered neural frame buffers can be combined with the ones produced by the standard rendering pipeline for resolving occlusions and alpha blending with the background. For example, the translucency of fur blends accurately with the floor or the background as the animal walks or jumps using our scheme.

\section{Neural Animal Motion Synthesis} \label{sec:mocap}

For ARTEMIS to provide convincing motion controls, e.g., to guide an animal to walk or jump from one location to another, it is essential to synthesize realistic intermediate motions as the animal moves. For human motions, data driven approaches now serve as the gold standard. However, a sufficiently large animal mocap dataset is largely missing. We hence first set out to acquire a new animal mocap dataset along with a companion deep animal pose estimation that we intend to share with research community. Next, we tailor a neural character animator ~\cite{starke2020local} for animals that maps the captured skeletal motions to abstract control commands and motion styles. 

\subsection{Animal Motion Capture}\label{sec:mocap setting}
\begin{figure}[t]
  \includegraphics[width=\linewidth]{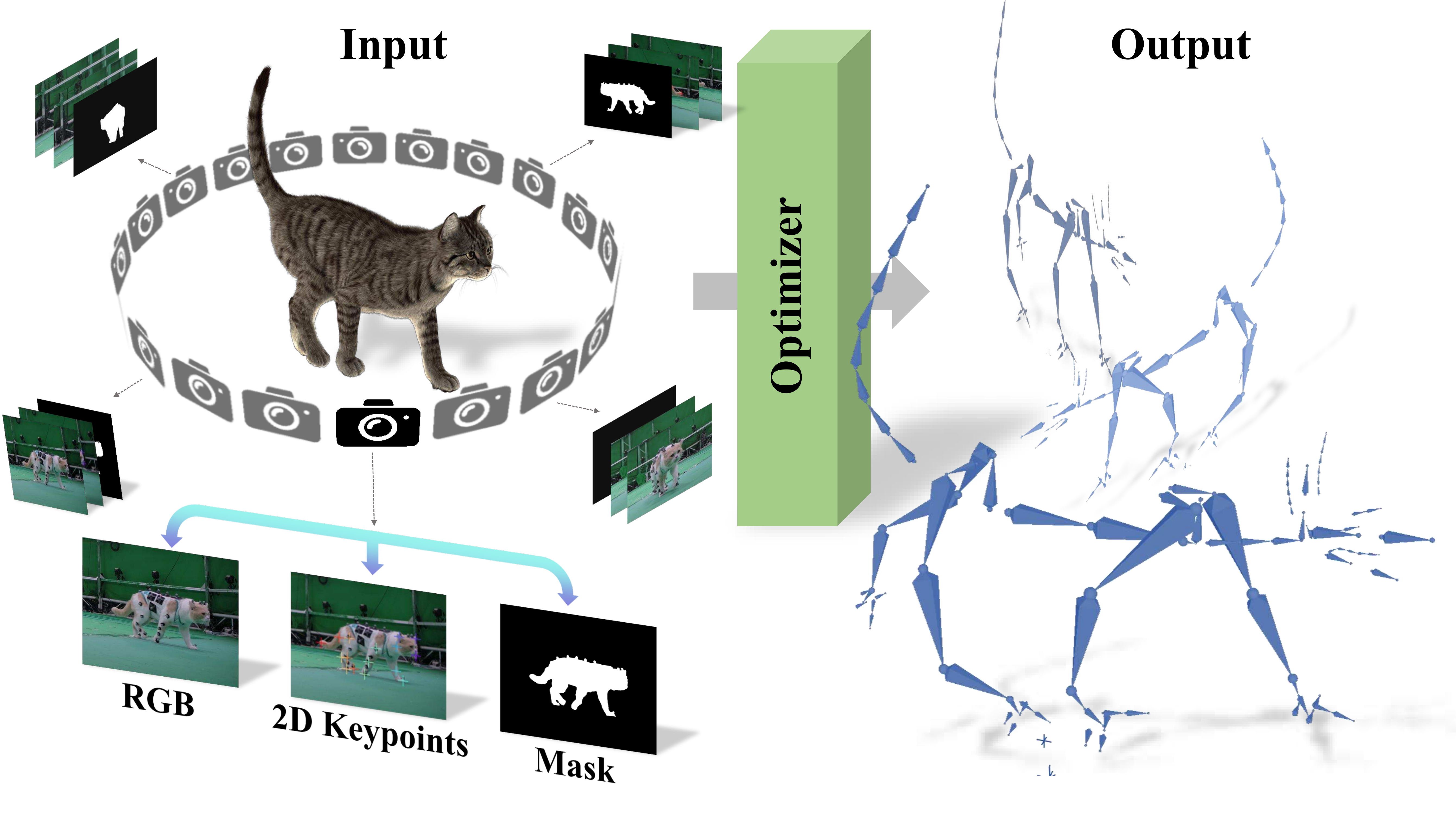}
  \caption{We estimate the animal skeleton, SMAL parameters more specifically, from the multi-view video data and Vicon MoCap data. We adopt an optimization framework to regress SMAL parameter from multiview observations of 2D keypoints and silhouettes and Vicon Observations.}
  \label{fig:mocap_pipeline}
\end{figure}

For data acquisition, we have constructed two types of animal motion capture systems, first composed of an array of RGB cameras for animals of large scales and the second combines RGB cameras and Vicon cameras for docile and small pets, as shown in Fig.~\ref{fig:mocap_pipeline}.

\label{sec:method}
\begin{figure}[t]
  \includegraphics[width=\linewidth]{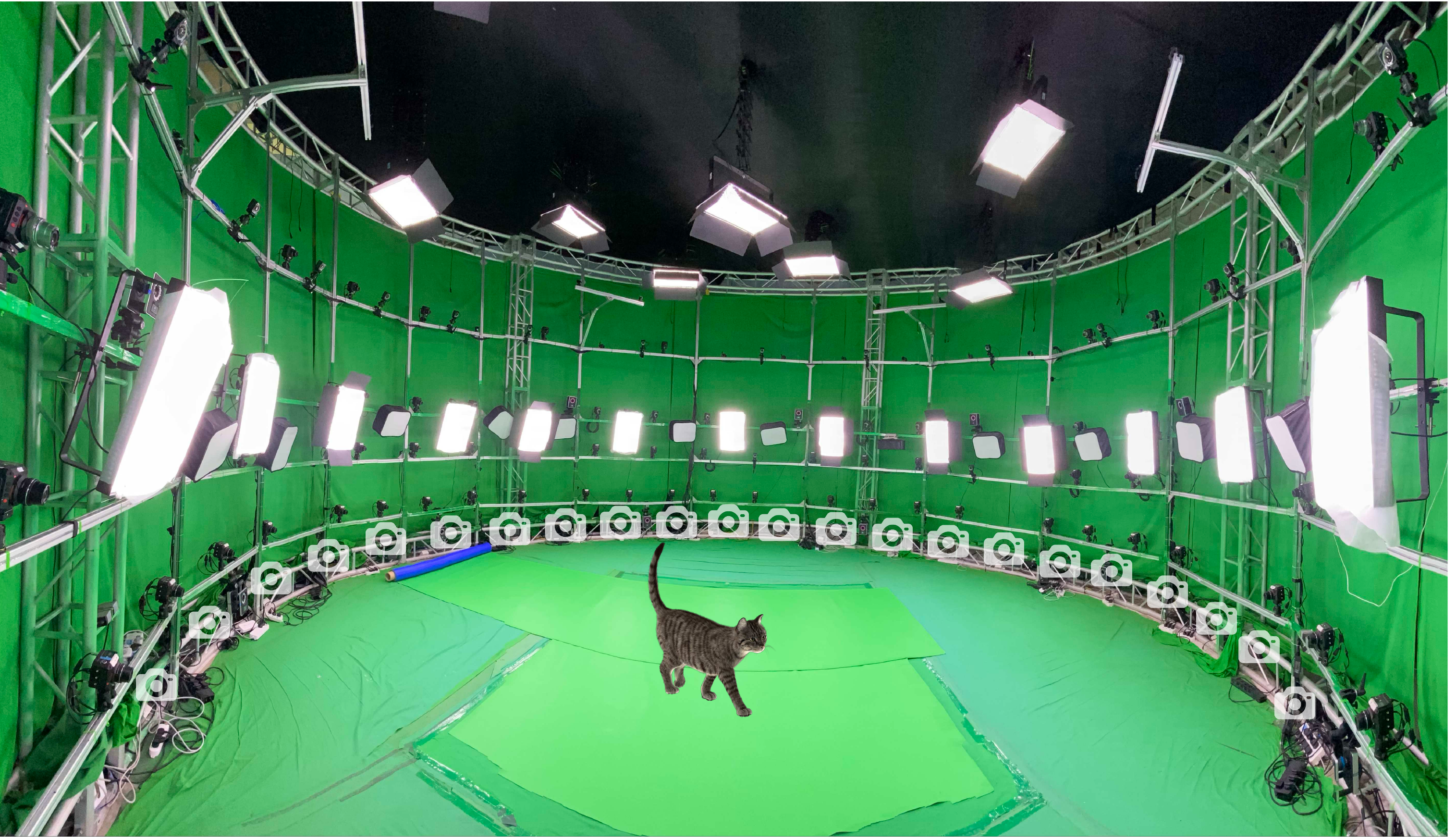}
  \caption{Setup of our animal motion capture system. The system consists of 12 Vicon cameras and 22 Z-CAM cinema cameras surrounding the animal. }
  \label{fig:mocap_setting}
\end{figure}

\paragraph{MoCap Processes.}
We observe that quadrupeds, despite having similar skeletal structures across species, exhibit drastically different shapes and scales. It is simply impossible to capture a mocap dataset suitable for all types of quadrupeds. We hence set out to learn motion priors from docile and small pets and transfer the priors to larger animals such as tigers and wolves. For the latter, we further enhance the prediction accuracy using a multi-view RGB dome in partnership with the zoo and circus.  

Fig.~\ref{fig:mocap_setting} shows our capture system for dogs and cats where we use 12 Vicon Vantage V16 cameras evenly distributed surrounding the animal subject, each capturing moving IR reflective markers at 120 Hz. We add additional 22 Z-CAM cinema cameras interleaved with Vicon and capturing at a $1920 \times 1080$ resolution at 30 fps. Cross-calibrations and synchronization are conducted in prior of actual motion capture. For small animals, we place the hybrid capture system closer to the ground so that they can clearly acquire limb movements while resolving occlusions between legs. We hire a professional trainer to guide these small animals to perform different movements including jumping and walking. Vicon produces highly accurate motion estimations that are used as priors for RGB camera based estimation.

For large animals, it is infeasible to place markers on large animals such as horses, tigers, elephants and etc. We partner with several zoos and circuses to deploy the RGB dome system where we use 22 \~ 60 cameras and adjust their heights and viewing directions, depending on the size of the performance area. It is worth mentioning that several circuses further provided us surveillance footages of their animals where we manually select the usable ones and label their poses.

\paragraph{2D Key-points and Silhouettes Estimation.}
To automatically extract skeletons on multi-view RGB images for animal motion inference, we first estimate the 2D key-points and silhouettes in images ~\cite{biggs2020left, zuffi2019three}. For key-points, we combine DeepLabCut ~\cite{MathisWarren2018speed} and the SMAL~\cite{zuffi20173d} model for processing imagery data towards quadrupeds, augmented with light manual annotations. Specifically, we annotate joints defined in the SMAL model on the first $10\%$ frames and then allow DeepLabCut to track and estimate 2D poses for remaining frames. For silhouette extractions, we directly use the off-the-shelf DeepLab2 ~\cite{weber2021deeplab2} with their pre-trained models.

\paragraph{Animal Pose Estimator.} We adopt the parametric SMAL animal pose model. To briefly reiterate, SMAL is represented as a function $M(\beta, \theta, \gamma)$, where $\beta$ is the shape parameter, $\theta$ the pose, and $\gamma$ the translation. $\Pi(x, C_j)$ denote the projection of a 3D point $x$ on the $j$-th camera whereas $\Pi(M, C_i)$ represents the projection of the SMAL model to camera $C_i$. Our goal is to recover SMAL parameters $\theta, \phi, \gamma$ from the observed 2D joints and silhouette. Assume all RGB and Vicon cameras in our system are calibrated. We extend the key point reprojection error $E_{kp}$, silhouette error $E_s$ as presented in \cite{zuffi20173d} to multiview inputs. In our setup, we assume known animal species and therefore append shape constraints $E_\beta$ from \cite{zuffi20173d} to enforce the distribution of $\beta$ to match the pre-trained species type. For our multi-view setup, we further add a 3D key-point constraint $E_{3d}$ and a mocap constraint $E_{m}$ as:

\begin{equation}
    E_{3d} (\Theta; V) = \sum_{k} \big \| V_{k} -\textbf{Tr}({v_{k}^i}) \big \|_2
\label{eq:3d_joint_loss}
\end{equation} 
\noindent where $V_k$ denotes the $k$th SMAL joint in 3D space and $v_{k}^i$ the observation of $V_k$ in camera $i$. $\textbf{Tr}(\cdot)$ corresponds to the triangulation operator. We hence have:
\begin{equation}
    E_m (\Theta; M) = \sum_{k} \| M_k - \frac{1}{ | \Omega |} \sum_{i \in \Omega_k} \mathcal{T}_k^i ( \textbf{Tr}({v_{k}^i} ) \|_2
\label{eq:marker_loss}
\end{equation}
\noindent where $M_k$ is position of the $k$th mocap marker. $\Omega_k$ corresponds to the set of joints used for recovering $M_k$. $\mathcal{T}_k^i$ is the transformation between the $k$th marker and vertex $i$. Notice not all vertices are usable for recovering $M_k$ because the animal motion is non-rigid. We thus find $\Omega_k$ by identifying the vertices that have correspond to constant $\mathcal{T}_k^i$ across the motion sequence. 

For human pose estimation, the SMPL-X human model~\cite{SMPLX2019} introduces a prior using Variational AutoEncoder~\cite{Kingma2014} to penalize on impossible poses. We adopt a similar idea: we per-train an animal motion prior using the same framework with all recovered skeletal poses of quadrupeds in our dataset and then use $E_{prior}$ to penalize on the deviations from the prior. We thus can formulate the animal pose estimator in terms of an optimization problem as:

\begin{equation}
\beta, \theta, \gamma \leftarrow \arg\min (E_{kp} + E_{s} + E_{3d} + E_{m} + E_{\beta} + E_{prior})
\label{eq:mocap_loss}
\end{equation} 
\noindent On RGB only captured images, we remove the marker loss term $E_{m}$. With recovered $\beta, \theta, \gamma$ for all frames, we can construct an animal motion sequence $ \{ \theta_j \}$ and $\{ \gamma_j \}$.

\subsection{Motion Synthesis} \label{sec:controller}

Our task is to generate virtual pets that exhibit realistic motions in response to a user's command. Physics-based controls are widely used in industry to generate vivid periodical motion patterns. However, those physical rules such as momentum conservation and reaction force can be confusing to users as they provide less direct guidance.
Hence, we leverage the recent advances in data-driven methods and human movement animation, i.e., Local Motion Phases (LMP)~\cite{starke2020local}, for our animal motion synthesis to learn movements from asynchronous behaviors of different body parts. Different from \cite{starke2019neural} that requires tedious manual annotation, LMP features can be automatically extracted and subsequently trained on unstructured motion data collected on animals.

\begin{figure}
  \includegraphics[width=\linewidth]{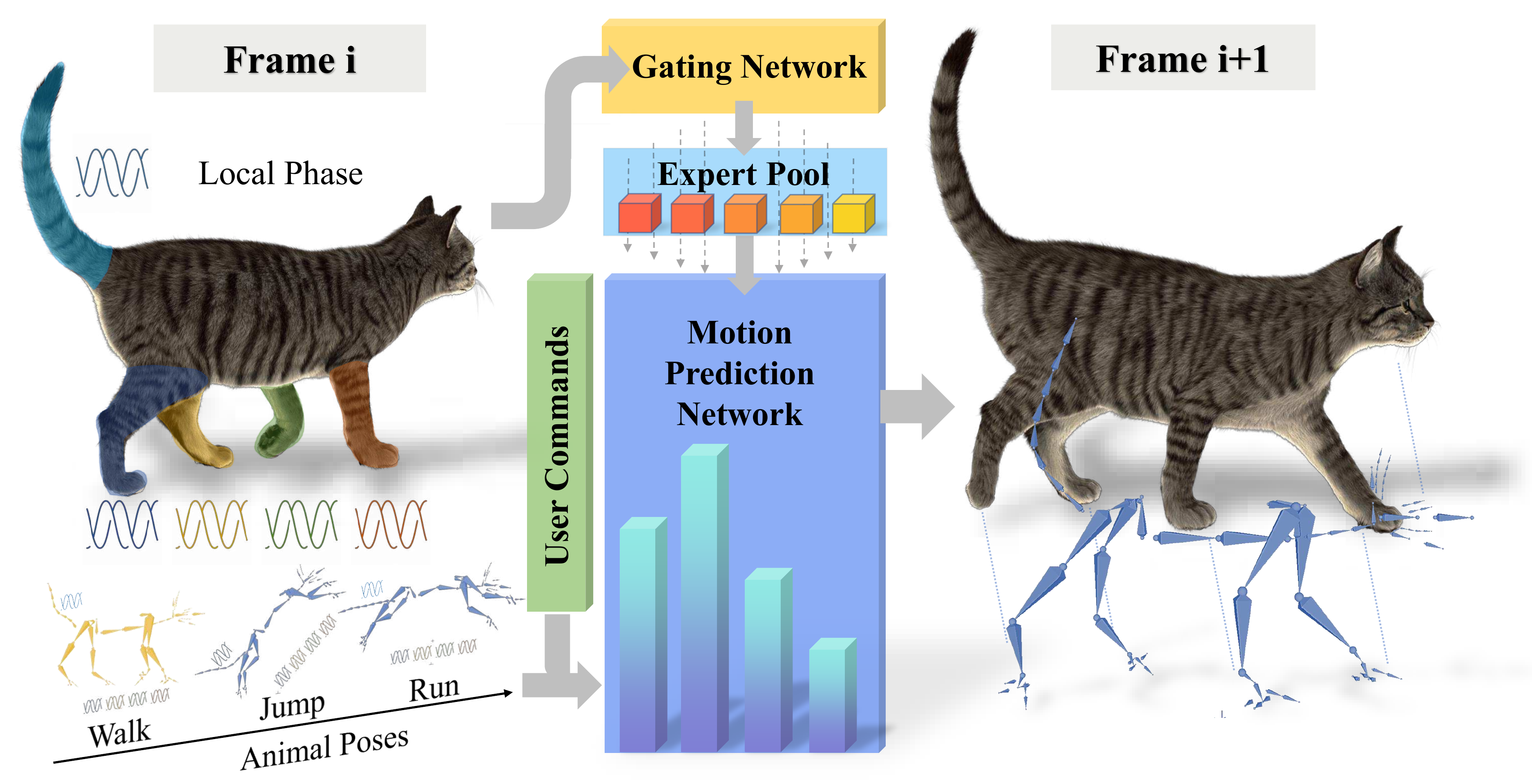}
  \caption{\textbf{Neural Animal Controller Pipeline.} Leveraging ideas of Local Motion Phases, the neural animal controller is composed of a gating network and a motion prediction network. The gating network inputs local phases and calculates the expert blending coefficients. The coefficients are used to generate the motion prediction network. Then, the user-given control signals and motion information from previous frames are sent into the motion prediction network, which uses them to predict the motion information of the next frame.}
  \label{fig:neural_state_machine}
\end{figure}

\paragraph{Controlled Motion Synthesis with Local Motion Phases.}
To enable interactions between the user and the virtual animal, it is critical to synthesize animal motions according to user commands, e.g., for accomplishing specific tasks. We adopt the LMP framework that consists of a gating network and a motion prediction network, as shown in Fig.~\ref{fig:neural_state_machine}. 
The gating network takes the motion states (joint locations, rotations, and velocities) of the current frame and the past frames as inputs and computes expert activation, which is then dynamically blended with motion states and sent to the motion prediction network.
Taking the expert weights and control signals from users, the motion prediction network would be able to calculate the motion state for future frames. The process can be expressed as:

\begin{equation}
    \mathcal{M}_i = \Phi (\mathcal{M}_{i-1}, c)
\end{equation}
where $\mathcal{M}_i$ is the motion state at frame $i$, $c$ is the user control signal. In our implementation, we define the set of control signals as \{`Idle', `Move', `Jump', `Sit', `Rest'\}.
Under these control signals, we extract the LMP feature accordingly, in terms of the contact state between the animal's feet or hips with the ground. The contact label is also extracted automatically by calculating the difference in positions and velocities between end-effectors and the environment collider. Furthermore, we process our motion data to calculate the action labels automatically. 

For example, we compute Idle and Move states by calculating the root velocities and mapping them into values from 0 to 1. We detect Jump, Sit and Rest labels by jointly checking the position, velocity, and contact information and mapping them to values between 0 and 1. Specifically, Jump labels have no contact and y-axis velocity, whereas the Sit and Rest Labels have no velocity and all end-effectors contact with the environment but differ in the position of the head of the animal.

\paragraph{Motion transfer to articulated model.}\label{sec:motion transfer}
Notice that the generated motion state from LMP is bound to the skeleton captured in our Mocap systems. Yet the base models used for our neural pets are created by artists, and their skeletons do not match with the captured animals. We need to transfer the motions generated by LMP to animal models created by artists while keeping the model shape. Therefore, for each type of virtual animal, we manually apply offsets constraining the rotation and translation components in the rest pose. We then apply forward-kinematics and inverse-kinematics to calculate the target motion state and use transformation limits to fine-tune the impossible states. With the transformed motion data, we can transfer motion states from one to another and drive the neural animals to move freely.

\begin{figure*}[thp]
  \includegraphics[width=1.0\linewidth]{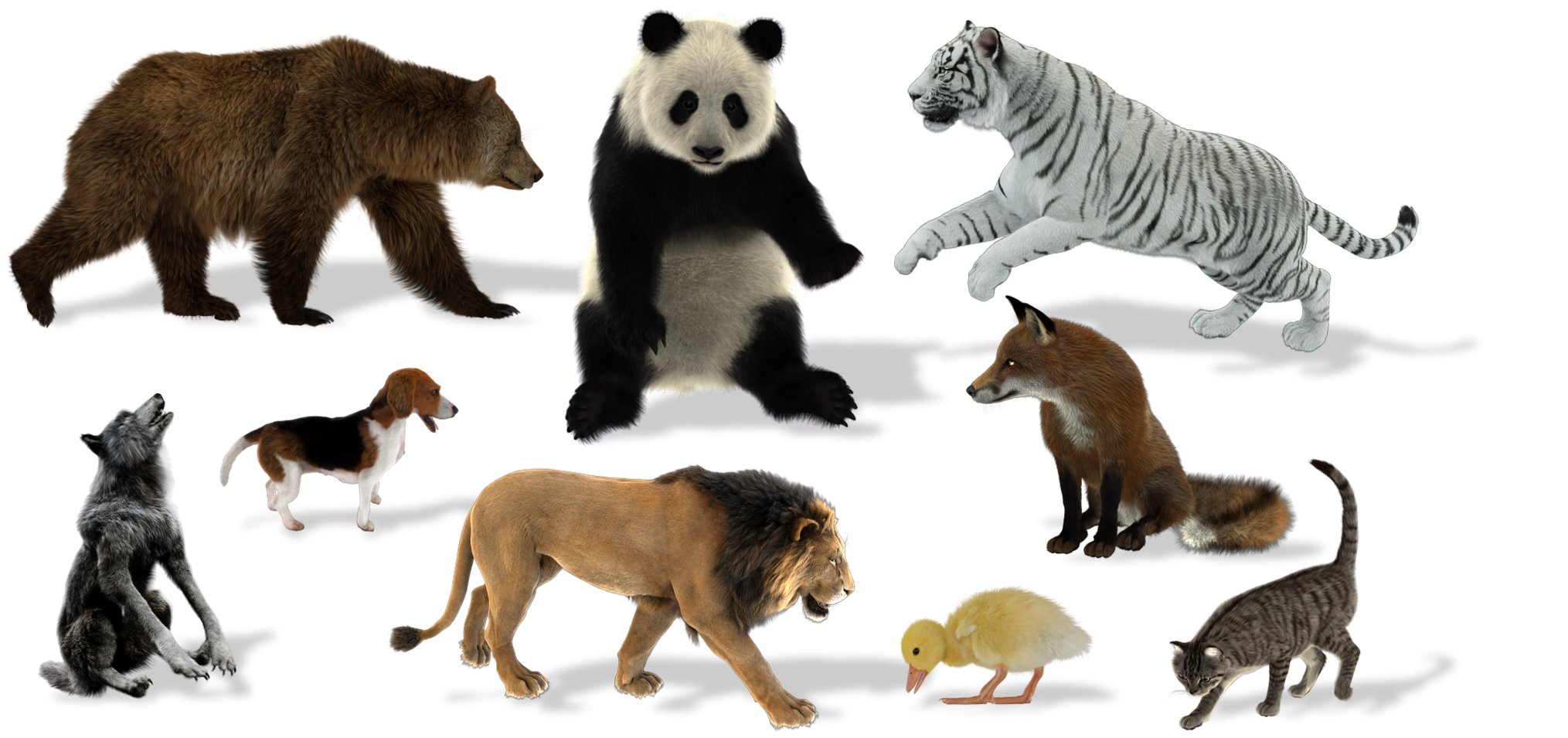}
  \caption{\textbf{Neural-generated animals in our ARTEMIS system.} We show our synthesized results of different neural volumetric animals in representative poses.}
  \label{fig:pet_gallery}
\end{figure*}

\section{Neural Animals in Immersive Environments}\label{sec:VR}
In previous sections, we have described how to train our animatable neural volumetric representation to synthesize animals under arbitrary poses and views, along with a hybrid rendering pipeline. Our neural motion controls and synthesis, enabled by our animal mocap data, provide convenient interfaces to guide the animal's movements. In this section, we assemble all these modules coherently under the VR setting as the ultimate ARTEMIS system, where a user can interact with the animal.

\paragraph{Interactions.}  \label{sec:mode}
Our motion synthesis module guides virtual animal movements by explicitly pointing to the target location and providing action types. We further explore high-level control patterns, emulating a commonly used set of commands by pet owners: 
\begin{itemize} 
	\setlength\itemsep{0em}
	\item Companion: As the user can move freely in virtual space, so will the virtual animal. It will follow the user. 

	\item Go to: The user points to a 3D location in virtual space, and the animal reaches the target destination automatically. In a complicated environment where obstacles are presented, we use the A-Star algorithm to find the path for the animal to follow. The user can also control the speed of the movement, i.e., the animal can either walk to or run/jump to the destination.  
	
	\item Go around a Circle: The user specifies a location, and the animal will reach the location and continue going around in circle. The user can even point themselves as the destination and ended being surrounded by single or multiple animals. 
	
	\item Size and Speed Adjustment: Since our neural animal representation, as a continuous implicit function, can support infinite resolution, the user can adjust the animal's size by simply adjusting its octree accordingly. The user can also adjust the movement speed, either slowing it down or speeding it up. In particular, slow motion provides a fancy visual effect where the user can observe the animal at a close distance as if time freezes.
	
	\item Free Mode: when no command is given, the animal can take any reasonable movements, such as exploring the virtual world themselves. In our implementation, we randomly set destinations within a time interval where the animal reaches the destination one after another. 
	
\end{itemize} 

\begin{figure*}[t]
	\includegraphics[width=1.0\linewidth]{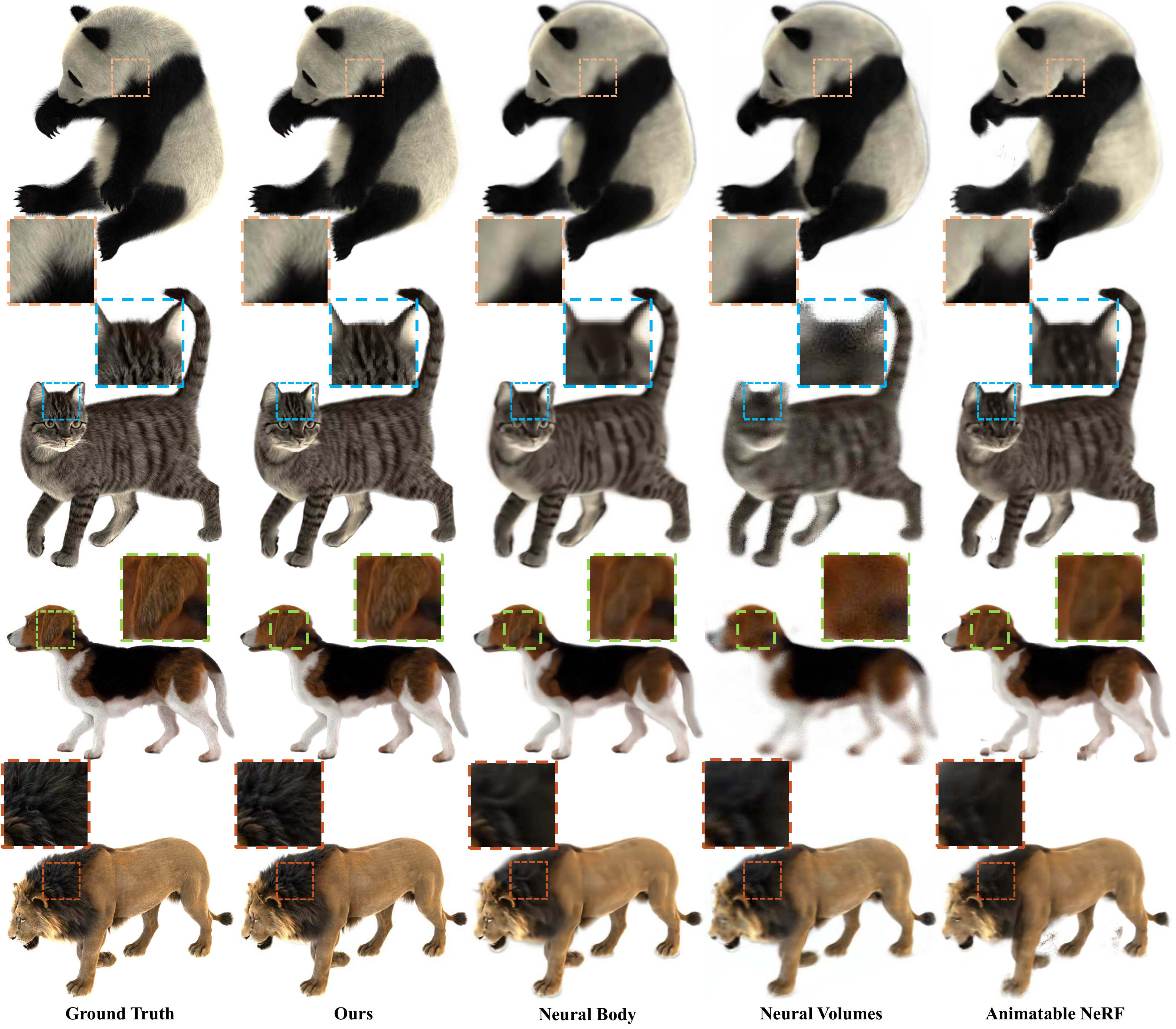}
	\caption{\textbf{Qualitative comparison for dynamic appearance rendering} against NeuralVolumes, NeuralBody and AnimatableNeRF. Note that our approach achieves more photo-realistic rendering with finer details for appearance and fur.}
	\label{fig:dynamic_comparison}
\end{figure*}

\paragraph{Implementations Details.}
The entire control flow for producing neural animals under the VR setting is as follows: We first obtain the pose of the VR headsets along with the controller status from OpenVR. Next, we then use LMP control signals and the current state to generate the estimated motion parameters. Based on the parameters, we rig the canonical model using LBS and then construct an octree for each frame. Finally, our neural rendering pipeline traces into the octree to generate the feature maps and subsequently conduct rendering where the background environments, obstacles in 3D environments, etc, are rendered using a standard graphics pipeline and fused with the neural rendering results. 

Note that for VR applications, it is critical to generate two consistent views for presenting to the left and right eyes of a user, respectively, in the VR headsets. Inconsistent views between eyes may cause discomforts, such as nausea and dizziness. Thus when we deploy the trained neural animals to our system, we adopt the stereo loss proposed in LookinGood~\cite{10.1145/3272127.3275099}, to finetune the trained model for better view consistency between left and right eyes in a self-supervised learning manner.

We follow the high cohesion and low coupling strategy to distribute time consumption as even as possible where the communication between core parts is lightweight via shared CUDA memory. The average computation time breakdown of individual parts are: ~10ms for handle input (e.g., update the left and right camera poses, handle the controller action, controller environment), ~29ms for motion synthesis (motion generation and motion transfer),  ~10ms for octree construction, ~25ms for stereo neural rendering and environment synthesis, ~10ms for output assembly and submit. By tailoring the assembly line of various components, we manage to produce the rendering at around 30 fps. The supplementary video shows live recordings of several examples.

\begin{figure*}[t]
	\includegraphics[width=1.0\linewidth]{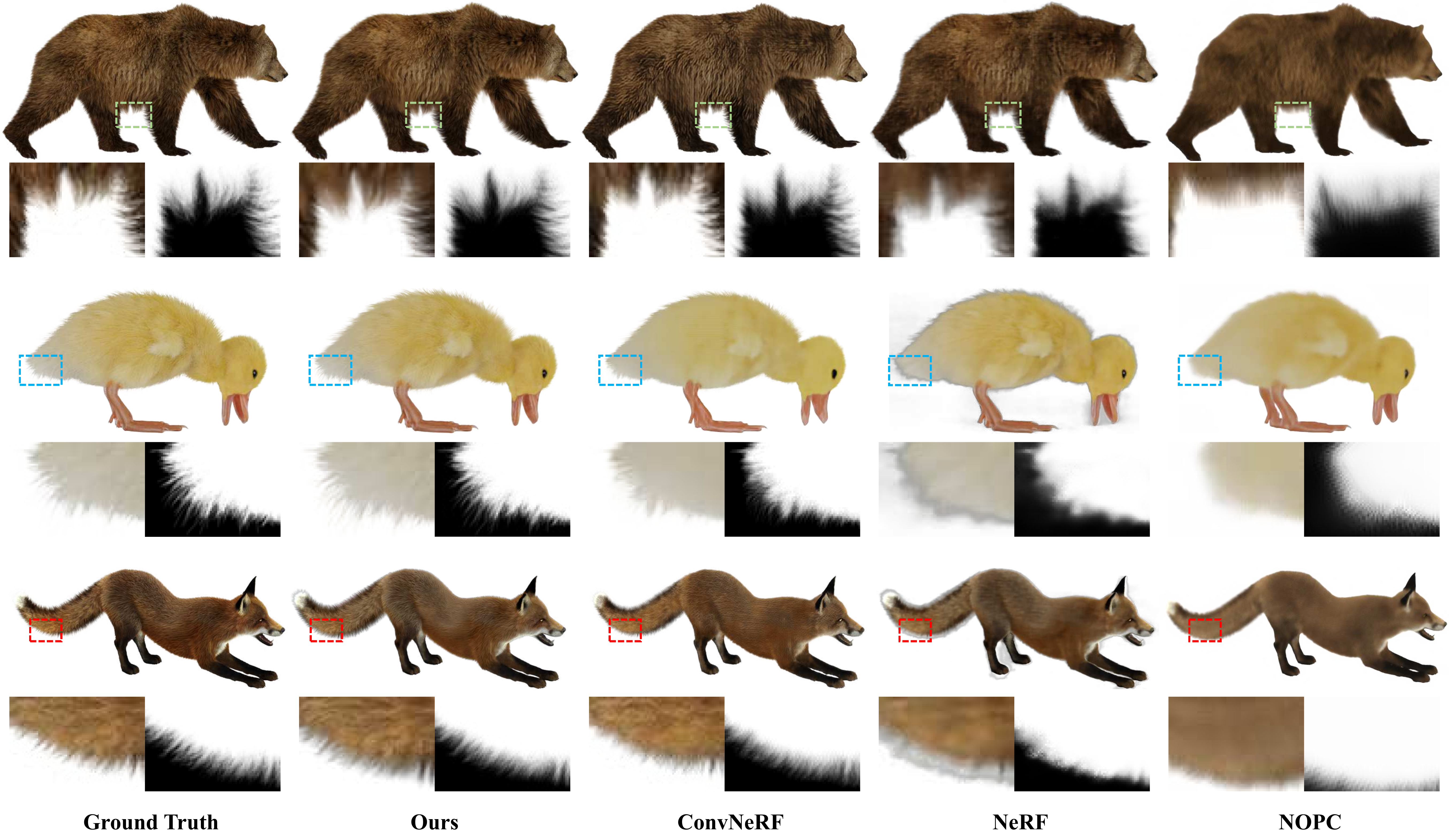}
	\caption{\textbf{Qualitative comparison for free-view appearance and opacity rendering of static scene} with NOPC, NeRF and ConvNeRF. Our approach achieves modeling high-frequency geometry details and generates almost the same RGBA images with the ground truth.}
	\label{fig:frame_compare}
\end{figure*}

\section{RESULTS}\label{sec:results}
In this section, we evaluate our ARTEMIS under various challenging scenarios. We first report our dynamic furry animal datasets, which are used for training our NGI animals, and our training and experimental details. Then, we provide the comparison of our neural rendering scheme with current state-of-the-art (SOTA) methods for both dynamic appearance rendering and static opacity generation, which demonstrate that our method better preserves high-frequency details. We also conduct a detailed runtime performance analysis of our rendering approach and provide extra evaluation for our motion capture scheme. We further provide the VR applications of our ARTEMIS using consumer-level VR headsets where a user can intimately interact with various virtual animals. Finally, the limitation and discussions regarding our approach are provided in the last subsection.

\paragraph{Dynamic Furry Animal Dataset}
To evaluate our ARTEMIS (neural pets) system, especially for our NGI animals, we seek help from traditional CGI animal animation and rendering pipelines to generate a comprehensive training dataset.
Specifically, our Dynamic Furry Animal (DFA) dataset contains nine high-quality CGI animals with fiber/strands based furs and skeletal rigs modeled through tedious artists' efforts, including panda, lion, cat, etc. 
We utilize the commercial rendering engine (e.g., MAYA) to render all these CGI animal characters into high-quality multi-view $1080\times1080$ RGBA videos under various representative skeletal motions.
Specifically, we adopted 36 camera views which are uniformly arranged around a circle towards the captured animal, and the number of representative poses ranges from 700 to 1000 for each animal.
Note that on average it takes about 10 minutes to render a single frame of our high-quality DFA dataset using commercial render farms.
For training an NGI animal from the corresponding CGI asset, we uniformly select 30 views to conduct per-animal training and leave six views as test data for evaluation.
We train our NGI animals at $960\times540$ image resolution using 4 Nvidia TITAN RTX GPUs, and it takes about two days for the training process.     
Fig.~\ref{fig:pet_gallery} demonstrates the representative rendering results of our NGI animals under various novel skeletal poses, where the details of appearance, fur, and opacity are faithfully reconstructed with real-time performance. 
Besides, our high-quality DFA dataset will be made publicly available to the community to stimulate future research about realistic animal modeling.

\subsection{Comparison to Dynamic Neural Rendering} \label{comparison}
Here we compare our neural rendering pipeline in ARTEMIS with recent SOTA methods for dynamic scene rendering. 
For thorough comparison, we compare against both the radiance field based methods, including NeuralBody~\cite{peng2021neural} and AnimatableNeRF~\cite{peng2021animatable}, as well as the volume-based approach NeuralVolumes~\cite{10.1145/3306346.3323020}.
Note that our NGI animal requires an additional pre-defined skeletal rig with skinning weights from a corresponding CGI digital asset.
Thus, for fair comparisons, we train the baseline models using the same experiment settings as our approach and adopt the ground truth mesh vertices and skinning weights of our CGI animal models to train NeuralBody and AnimatableNeRF. 

Fig.~\ref{fig:dynamic_comparison} provides the qualitative comparison results of our approach against the above baselines.
Note that only fur with color much different from the background can be recognized as fur, like lion manes and cat forehead. 
Non-surprisingly, the three baseline methods suffer from artifacts of blurry fur and loss of appearance details.
Specifically, NeuralVolumes suffers from severe noises and blurring effects due to its limited modeling ability, which is most obviously on the cat forehead. 
NeuralBody can only generate low-frequency fur details as blurry effects, while AnimatableNeRF behaves slightly better but still cannot faithfully recover the furry details.
In stark contrast, our approach generates much more clear details favorably, especially for those furry regions which are common in animals.
For quantitative evaluation, we adopt the peak signal-to-noise ratio (PSNR), structural similarity index (SSIM), and Learned Perceptual Image Patch Similarity (LPIPS) as metrics to evaluate our rendering accuracy.
As shown in Tab.~\ref{table:dynamic_comparison}, our approach significantly outperforms the other baselines for all the above metrics, demonstrating the superiority of our approach in preserving rendering details.

\begin{table}[t]
	\caption{Quantitative comparisons of synthesized appearance images on different dynamic animals. Compared with NeuralVolumes, NeuralBody, and AnimableNeRF, our method achieves the best performance in all metrics.}
	\centering
	\resizebox{1\linewidth}{!}{
		\begin{tabular}{m{3em}<{\centering} m{4em}<{\raggedright } c c c c} 
			\toprule
			\makecell[c]{Method} & & \makecell[c]{Neural\\Body} & \makecell[c]{Neural\\Volumes} & \makecell[c]{Animatable\\NeRF} & \makecell[c]{ Ours} \\ %
			\midrule
			\multirow{3}{*}{\bf Panda}           & $\uparrow$ PSNR     & 30.38 & 30.11 & 26.51 & \textbf{33.63} \\ 
			& $\uparrow$ SSIM     & 0.970 & 0.965 & 0.957 & \textbf{0.985} \\
			& $\downarrow$ LPIPS  & 0.110 & 0.116 & 0.112 & \textbf{0.031} \\
			\midrule
			\multirow{3}{*}{\bf Cat}             & $\uparrow$ PSNR     & 30.77 & 28.14 & 31.37 & \textbf{37.54} \\ 
			& $\uparrow$ SSIM     & 0.972 & 0.951 & 0.973 & \textbf{0.989} \\
			& $\downarrow$ LPIPS  & 0.067 & 0.087 & 0.061 & \textbf{0.012} \\
			\midrule
			\multirow{3}{*}{\bf Dog}             & $\uparrow$ PSNR     & 32.37 & 26.80 & 31.19 & \textbf{38.95} \\ 
			& $\uparrow$ SSIM     & 0.978 & 0.945 & 0.975 & \textbf{0.989} \\
			& $\downarrow$ LPIPS  & 0.075 & 0.129 & 0.074 & \textbf{0.022} \\
			\midrule
			\multirow{3}{*}{\bf Lion}            & $\uparrow$ PSNR     & 30.11 & 29.59 & 27.87 & \textbf{33.09} \\ 
			& $\uparrow$ SSIM     & 0.956 & 0.947 & 0.944 & \textbf{0.966} \\
			& $\downarrow$ LPIPS  & 0.111 & 0.123 & 0.123 & \textbf{0.035} \\
			
			\bottomrule
		\end{tabular}
	}
	\label{table:dynamic_comparison}
\end{table}

\begin{table}[!t]
\caption{Quantitative comparisons of appearance and alpha on a single representative frame of different animals. Our approach achieves the best performance in almost all alpha-related metrics and comparable performance for RGB texture against ConvNeRF.}
\centering
\resizebox{1\linewidth}{!}{
	\begin{tabular}{m{3em}<{\centering} m{4em}<{\raggedright } c c c c} 
		\toprule
		\textbf{RGB} & & NOPC & NeRF & ConvNeRF & Ours \\ %
		\midrule
		\multirow{3}{*}{\bf Bear}            & $\uparrow$ PSNR     & 18.43 & 28.12 & \textbf{32.34} & 30.95 \\ 
		& $\uparrow$ SSIM     & 0.886 & 0.954 & 0.953 & \textbf{0.967} \\
		& $\downarrow$ LPIPS  & 0.140 & 0.113 & 0.063 & \textbf{0.038} \\
		\midrule
		\multirow{3}{*}{\bf Duck}            & $\uparrow$ PSNR     & 25.45 & 30.35 & 34.31 & \textbf{37.14} \\ 
		& $\uparrow$ SSIM     & 0.967 & 0.978 & 0.985 & \textbf{0.986} \\
		& $\downarrow$ LPIPS  & 0.075 & 0.091 & 0.052 & \textbf{0.026} \\
		\midrule
		\multirow{3}{*}{\bf Fox}             & $\uparrow$ PSNR     & 17.42 & 27.53 & \textbf{33.42} & 30.94 \\ 
		& $\uparrow$ SSIM     & 0.914 & 0.966 & 0.973 & \textbf{0.976} \\
		& $\downarrow$ LPIPS  & 0.106 & 0.099 & 0.047 & \textbf{0.029} \\
		\midrule
		\textbf{Alpha} & & NOPC & NeRF & ConvNeRF & Ours \\ [0.5ex]
		\midrule
		\multirow{3}{*}{\bf Bear}            & $\uparrow$ PSNR     & 17.89 & 31.65 & 36.37 & \textbf{40.13} \\ 
		& $\uparrow$ SSIM     & 0.918 & 0.986 & 0.992 & \textbf{0.995} \\
		& $\downarrow$ SAD    & 144.2 & 199.2 & 11.80 & \textbf{8.072} \\
		\midrule
		\multirow{3}{*}{\bf Duck}            & $\uparrow$ PSNR     & 19.77 & 30.09 & 33.02 & \textbf{36.81} \\ 
		& $\uparrow$ SSIM     & 0.849 & 0.923 & 0.990 & \textbf{0.994} \\
		& $\downarrow$ SAD    & 110.6 & 36.17 & 12.76 & \textbf{8.558} \\
		\midrule
		\multirow{3}{*}{\bf Fox}             & $\uparrow$ PSNR     & 15.68 & 23.81 & 34.90 & \textbf{36.43} \\ 
		& $\uparrow$ SSIM     & 0.903 & 0.968 & 0.993 & \textbf{0.995} \\
		& $\downarrow$ SAD    & 192.4 & 52.32 & 11.30 & \textbf{9.555} \\
		\bottomrule
	\end{tabular}
}
\label{table:frame_comparison}
\end{table}

\subsection{Comparison to Static RGBA Rendering} \label{comparison2}
Here we evaluate our approach for generating free-view opacity maps, which is important for fur modeling.
To the best of our knowledge, there are no previous works that can generate opacity maps for dynamic scenes.
Thus, we compare our approach against the recent SOTA methods on the opacity rendering task for static scenes.
For thorough comparison, we select three baselines, including the explicit point-cloud based method called Neural Opacity Point Clouds (NOPC)~\cite{Wang2020NOPC}, as well as the radiance field based methods NeRF~\cite{mildenhall2020nerf} and ConvNeRF~\cite{9466273}.

Fig.~\ref{fig:frame_compare} shows several RGB and alpha results of our method and other static methods.
\textit{NOPC} suffers from aliasing and ghosting on the boundary regions due to the interpolation of point-cloud based features.
\textit{NeRF} suffers from severe blur artifacts because of the limited representation ability of its MLP network, especially for high-frequency details, while \textit{ConvNeRF} improves the alpha and RGB details but still causes grid-like artifacts due to patch-based training strategy on limited sparse training views. In contrast, our approach achieves the best performance where we manage to compensate for missing views at a specific frame using views from other frames.

Differently, our approach generates more realistic opacity rendering and even supports dynamic scenes in real-time.
The corresponding quantitative result is provide in Tab.~\ref{table:frame_comparison}.
For the evaluation of the alpha map, we further adopt Sum of Absolute Distance (SAD) as metrics besides PSNR and SSIM.
Our approach outperforms all the baselines for alpha-related metrics and maintains comparable performance for RGB texture rendering against ConvNeRF. 
All these evaluations illustrate the effectiveness of our approach for high-quality fur and opacity detail rendering.

\begin{figure*}[t]
	\includegraphics[width=1.0\linewidth]{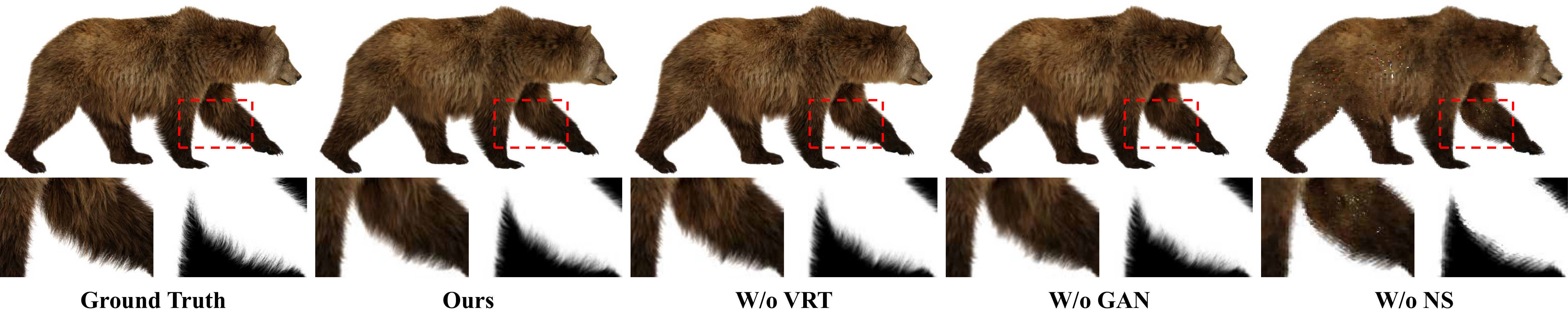}
	\caption{\textbf{Visualization of the ablation study} on w/o the NS, GAN loss and VRT loss modules (corresponding to the Tab.~\ref{table:ablation}).}
	\label{fig:ablation}
\end{figure*}

\subsection{Ablation Study} \label{ablation}
We conduct extensive ablation studies on the ``Bear'' data.
Let \textbf{w/o VRT} and \textbf{w/o GAN} denote our models trained without voxel regularization term (VRT) and GAN loss. Fig.~\ref{fig:ablation} shows GAN loss helps to preserve more high-frequency details and synthesized sharper alpha map, and VRT leads to slightly smoothed texture while maintaining temporal consistency. We further train our model to directly render RGBA images through volumetric without neural shading network (NS) (\textbf{w/o NS}). It suffers from noises and artifacts around the legs and abdomen due to geometry misalignment and limited volume resolution. The qualitative result in Tab.~\ref{table:ablation} well supports the contribution of each module.

\begin{table}[t!]

\begin{center}
\caption{Quantitative evaluation of the ablation studies on w/o the NS, GAN loss and VRT loss modules. Our full model achieves the best performance.}
\resizebox{1\linewidth}{!}{
\begin{tabular}{cccccccccc}
\toprule
\multicolumn{2}{c}{\multirow{2}{*}{Models}} & \multicolumn{3}{c}{$RGB$} & \multicolumn{3}{c}{Alpha} \\\cline{3-8}
\specialrule{0em}{1pt}{1pt}
\multicolumn{2}{c}{}                 & PSNR $\uparrow$ & SSIM $\uparrow$ & LPIPS $\downarrow$ & PSNR $\uparrow$ & SSIM $\uparrow$ & SAD $\downarrow$ \\
\midrule

\multicolumn{2}{l}{(a) w/o NS}       & 29.47 & 0.939 & 0.152 & 29.27 & 0.981 & 24.68 \\
\multicolumn{2}{l}{(b) w/o GAN}      & 34.29 & 0.964 & 0.051 & 35.84 & 0.991 & 12.47 \\
\multicolumn{2}{l}{(c) w/o VRT}      & 33.81 & 0.961 & \textbf{0.044} & 35.74 & \textbf{0.992} & 12.23 \\
\multicolumn{2}{l}{(d) ours}         & \textbf{34.43} & \textbf{0.965} & 0.045 & \textbf{36.18} & \textbf{0.992} & \textbf{11.92} \\

\bottomrule
\end{tabular}
}
\label{table:ablation}
\end{center}
\end{table}

\subsection{Runtime Performance Evaluation} \label{evaluation1}

We run our method on an Nvidia TITAN RTX GPU. Tab.~\ref{table:runtime_compare} compares the running time of our method for one frame with others. 
For dynamic scenes, we compare our full runtime, i.e., warping to target pose and rendering, with other baselines. Our method achieves real-time novel pose synthesis and rendering. 
For static scene, we compare our free-view rendering runtime.
Our method is much faster (in order of 10 times) than other methods, especially achieves three to four orders of magnitude rendering speed up to traditional CGI animal character, which further supports our novel ARTEMIS system.

\paragraph{Runtime Analysis.}
We further analyze the runtime of each part and the performance-speed trade-off on a single Nvidia GeForce RTX3090 GPU. 
As shown in Tab.~\ref{tab: runtime}, for a general case, where the resolution of our sparse volume is 512, animating (warp + build octree) costs around 10ms, volume rendering costs around 5ms, and the neural shading network costs around 13ms with half floating-point
precision. It costs 27.43ms in total. To further accelerate, we design a light shading network by removing the opacity branch and adding an output channel to the appearance branch to predict the opacity map, denoted as \textbf{Light}. The rendering time reduces to 24.84ms with a slight performance downgrade. Finally, by combining the light network with 256 volume resolution, denoted as $\textbf{256 + Light}$, it takes 22.69ms with a slight performance drop. We adopt this scheme to provide a better experience in our ARTEMIS system.

\begin{table}[t]
\caption{Runtime comparisons different Methods. Our method is significantly faster than existing methods and enables real-time applications.}
\centering
\resizebox{1.0\linewidth}{!}{
	\begin{tabular}{m{6em}<{\centering} c c c c c} 
		\toprule
		\textbf{Dynamic} & \makecell[c]{CGI} & \makecell[c]{Neural\\Body} & \makecell[c]{Neural\\Volumes} & \makecell[c]{Animatable\\NeRF} & \makecell[c]{ Ours} \\ %
		\midrule
		runtime (ms)    & $\sim 5\times10^5$ & 2353 & 181.7 & 18142 & \textbf{34.29} \\ 
		fps        & $-$ & 0.425 & 5.504 & 0.055 & \textbf{29.16} \\
		\midrule
		\textbf{Static} & \makecell[c]{CGI} & NOPC & NeRF & ConvNeRF & Ours \\ [0.5ex]
		\midrule
		runtime (ms)   & $\sim 5\times10^5$ & 51.23 & 18329 & 2599 & \textbf{20.38} \\ 
		fps        & $-$ & 19.52 & 0.055 & 0.385 & \textbf{49.07} \\
		\bottomrule
	\end{tabular}
}
\label{table:runtime_compare}
\end{table}

\begin{table}[t]
\caption{Runtime (in milliseconds) analysis of components of our model and performance-speed trade-off.}
\begin{center}
	\centering
		\begin{tabular}{r|ccc}
			\toprule
			Model        & Normal & Light & 256 + Light \\ \hline
			PSNR $\uparrow$ & \textbf{33.01}       & 32.80      &     32.47        \\
			SSIM $\uparrow$  & \textbf{0.991}       & 0.990      &      0.988       \\
			LPIPS $\downarrow$ & \textbf{0.010}    & 0.011     &      0.012        \\ \hline
			warp $\downarrow$    &  1.950       &   1.975    &    \textbf{1.915}         \\
			build otree $\downarrow$ &  7.990      &   8.321    &  \textbf{6.285}           \\ 
			volume render $\downarrow$     &   4.521     & 4.520      &  \textbf{2.478}           \\
			neural render $\downarrow$         &  12.97      & 10.03      &  \textbf{10.01}           \\ \hline
			total $\downarrow$            &   27.43     &  24.84     &   \textbf{22.69}         \\
			\bottomrule
		\end{tabular}
	\label{tab: runtime}
\end{center}
\vspace{-2mm}
\end{table}

\begin{figure*}[thp]
\includegraphics[width=0.9\linewidth]{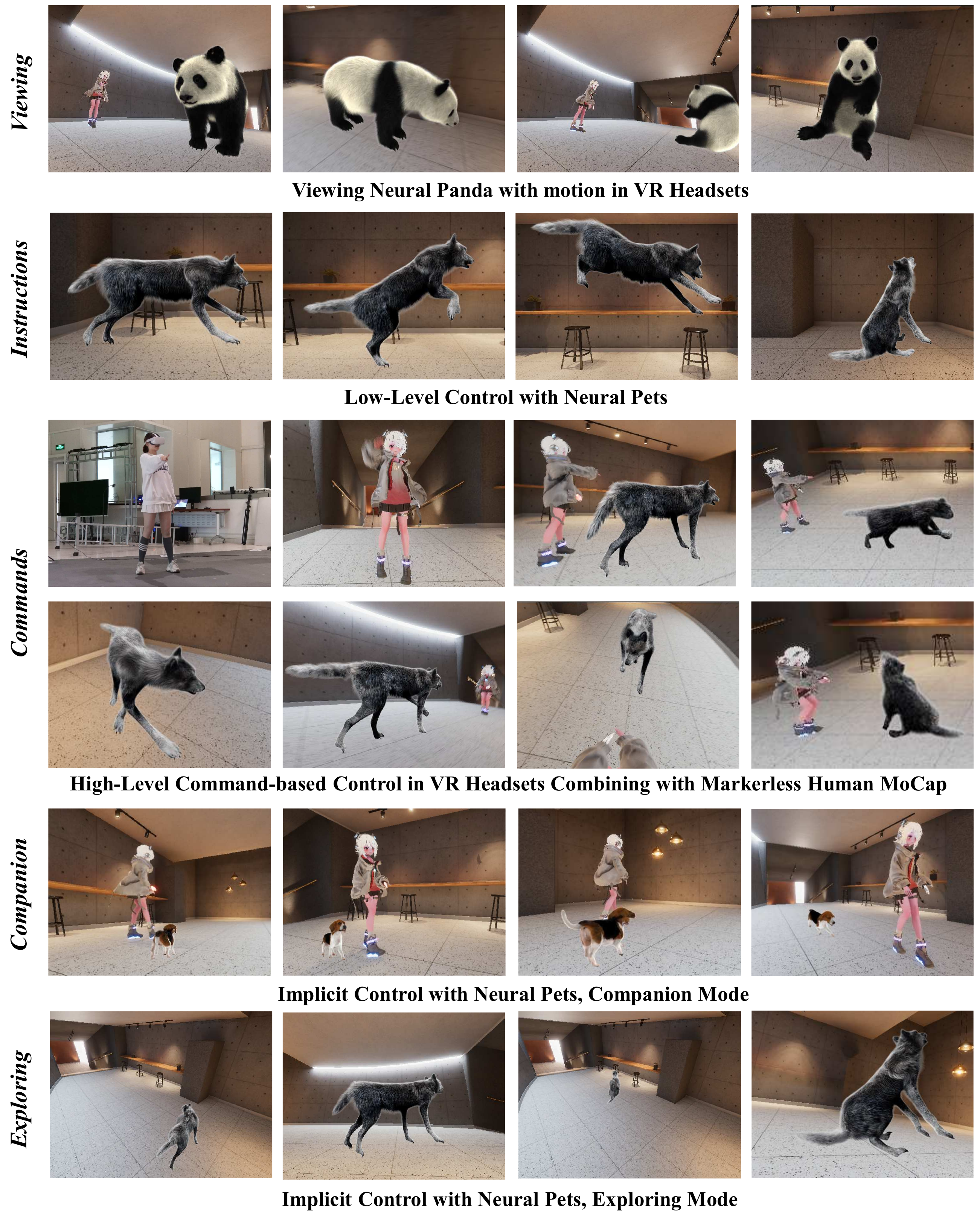}

\caption{\textbf{Example interactive modes of our ARTEMIS.} From top to bottom are different levels of interactions: viewing - we can observe the animal and see the fur clearly; instructions - drive the animals with explicit signals, such as `jump' or `sit'; commands - high-level control patterns, such as user points at a 3D location and the animal moves to there; companion - the animal follows the user; `exploration' - animals show random movements.}
\label{fig:VR-Gallery}
\end{figure*}

\begin{figure*}[thp]
\includegraphics[width=\linewidth]{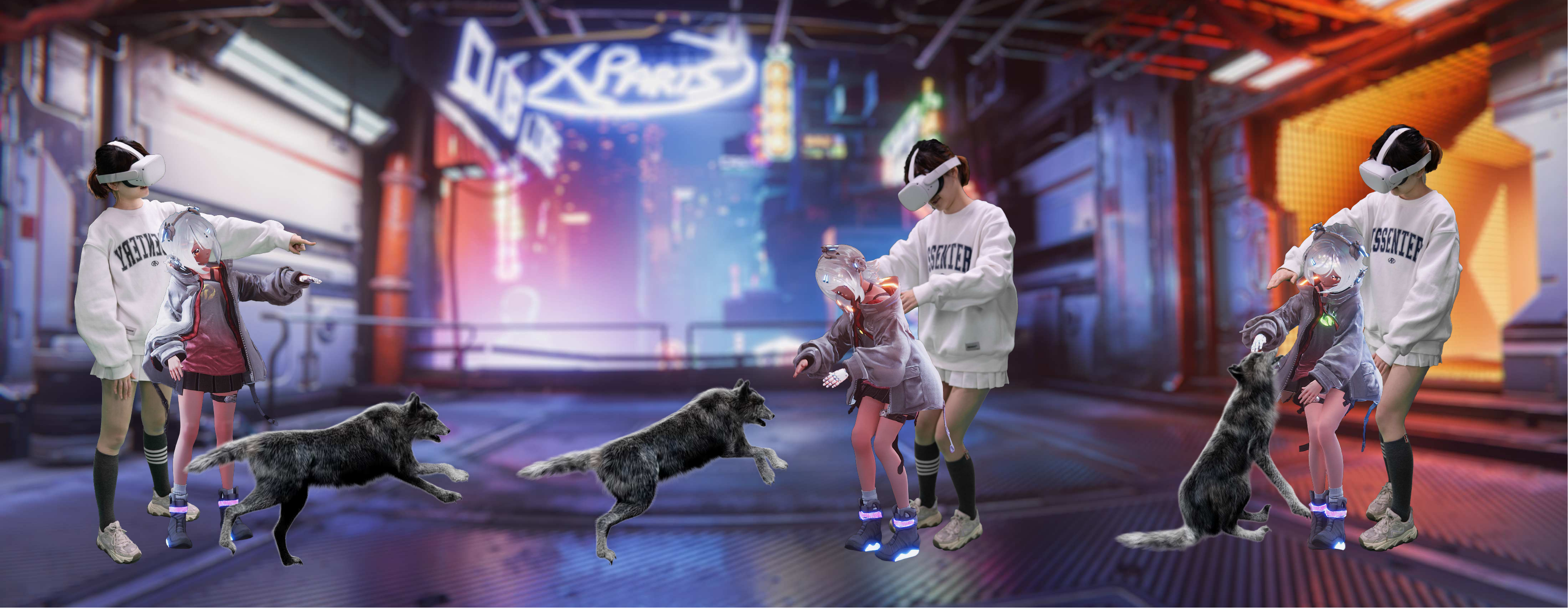}
\caption{\textbf{A concept demo of animal and user interaction in a VR environment.} We develop a VR demo to exhibit how users can interact with a virtual pet in a VR environment. We obtain the coordinate of the user in VR according to the controller, the wolf pet then plays with the user, e.g., jumping around the user. }
\label{fig:VR}
\end{figure*}

\subsection{Interactive NGI Animals in VR.}
As shown in  Fig.~\ref{fig:VR-Gallery}, we exhibit our rendering results in VR applications, including different level interactions and different perspectives. The first line, 'viewing' shows the vivid virtual panda we observed from the third and first view(VR headset), respectively. We can clearly see good fur effects even in VR headsets. The second line, 'Instructions' shows the low-level control animation results. We can explicitly drive our neural pets using control signals like 'Jump', 'Move' and 'Sit'. 'Commands' illustrates our high-level control pattern 'Go to'. The user points to a 3D location in virtual space, and the wolf reaches the target destination automatically. We can also call back the wolf by waving. Note that the speed of the movement is also controlled by the user. In 'Companion', our virtual pet will follow and accompany the user like a real pet. Finally, 'Exploring' shows our free mode, when no command is given, the animal can make any reasonable movements, exploring the virtual world themselves. 

We show more interactive results in Fig.~\ref{fig:VR}.

\subsection{Limitations and Discussions}
We have demonstrated the compelling capability of interactive motion control, real-time animation, and photo-realistic rendering of neural-generated (NGI) furry animals even with modern commercial-level VR headsets. However, as a first trial to bring photo-realistic NGI animals into our daily immersive applications with us humans, our ARTEMIS system still owns limitations as follows.

First, our NGI animal still heavily relies on a pre-defined skeletal rig and skinning weights of the corresponding CGI animal character. Despite the considerable improvement in terms of rendering time speed up from CGI to NGI animals, the problem of fully automatic neural animal generation from captured data without the artist's effort remains unsolved. In future work, we plan to combine the non-rigid tracking technique~\cite{unstructuredFusion} to alleviate our reliance on skeletal parameters.
Moreover, within the paradigm of neural rendering, our approach suffers from appearance artifacts for those unobserved animal body regions during training and under those challenging rendering views that are too different compared to the captured camera views. It is promising to combine the appearance prior of certain kinds of animals to alleviate such neural rendering shortcomings.
Our current rendering results still suffer from flickering artifacts when the rendering view or the captured animal is moving, even using our voxel regularization. We suspect that it's partially caused by the voxel collision of our dynamic Octree reconstruction. Besides, our technique cannot yet handle self-contact between distant body parts. In future work, we plan to explore a more elegant Octree indexing scheme and employ more temporal regularization to handle the flickering.
Moreover, our current NGI animal can only generate results within the baked lighting environment while rendering the CGI asset and thus cannot quickly adopt the lighting in a new environment compared to traditional rendering engines. It is interesting to enable more explicit lighting control on top of the current NGI animal design.

From ARTEMIS system side, our current interactions with NGI animals are still limited by the heavy reliance on human motion capture and pre-defined rule-based strategies. 
Fundamentally, current ARTEMIS design is based on only several relatively basic interactive modes.
Despite the unprecedented interaction experience of ARTEMIS, a more advanced and knowledge-based strategy for the interaction between NGI animals and us humans are in urgent need.
Moreover, more engineering effort is needed to provide a more immersive experience for current VR applications, especially for supporting the rendering of multiple animals.

\paragraph{Discussion}
ARTEMIS has the potential to create the neural animal directly from captured real animal data if some obstacles are conquered: it requires high-quality multi-view videos of furry animals with rich fur details while obtaining such videos of real animals remains difficult. Animals occupy a much smaller portion within the images than human performers, and therefore their captured videos are generally of a low resolution. Further, balancing the exposure setting (aperture vs. shutter vs. gain) to capture fast motion and shape images simultaneously can be very tricky. In contrast, skeleton extraction is generally robust if the camera setting is dense, as shown in the paper. Recent excellent work BANmo~\cite{yang2021banmo} also provides promising insights for animals from casual videos.

On the other hand, compared to rasterization-based schemes for fast rendering hair and fur (e.g., EEVEE), our work, instead, aims to show that neural rendering provides a promising alternative that can tackle shape deformations similar to traditional meshes but implicitly. In addition, such neural representations can potentially handle real animations without manual model building: by building a neural representation from a multi-view video sequence of a real animal, one can potentially conduct photo-realistic rendering of the animal under unseen poses without explicitly modeling the underlying geometry, let alone fur. Besides, furry objects are challenging to model using meshes: it requires extensive labor by artists and the use of ultra-dense meshes where implicit representation such as NeRF and its extensions can relieve artists from such labor.

\section{CONCLUSION}
We have presented a neural modeling and rendering pipeline called ARTEMIS for generating articulated neural pets with appearance and motion synthesis.
Our ARTEMIS system has interactive motion control, realistic animation, and high-quality rendering of furry animals, all in real-time.
At the core of our ARTEMIS, our neural-generated (NGI) animals renew the rendering process of traditional CGI animal characters, which can generate real-time photo-realistic rendering results with rich details of appearance, fur, and opacity with notably three to four orders of magnitude speed-up.
Our novel dynamic neural representation with a tailored neural rendering scheme enables highly efficient and effective dynamic modeling for our NGI animal. 
Our robust motion capture scheme for real animals, together with the recent neural character control scheme, provides the controllable ability to interact with our NGI animals with more realistic movements.  
Moreover, our hybrid rendering engine enables the integration of ARTEMIS into existing consumer-level VR headset platforms so as to provide a surreal and immersive experience for users to intimately interact with virtual animals as if in the real world.
Extensive experimental results and VR showcases demonstrate the effectiveness of our approach for neural animal modeling and rendering, supporting immersive interactions unseen before.
We believe that our approach renews the way humans perceive and interact with virtual animals, with more immersive, realistic, and responsive interaction experiences, with many potential applications for animal digitization and protection or fancy human-animal interactions in VR/AR, gaming, or entertainment.

\begin{acks}
The authors would like to thank Junyu Zhou and Ya Gao from DGene Digital Technology Co., Ltd. for processing the CGI animals models and motion capture data. Besides, we thank Zhenxiao Yu and Heyang Li from ShanghaiTech University for producing a supplementary video and figures.

This work was supported by NSFC programs (61976138, 61977047), the National Key Research and Development Program (2018YFB2100\\500), STCSM (2015F0203-000-06) and SHMEC (2019-01-07-00-01-E00003).

\end{acks}

\bibliographystyle{ACM-Reference-Format}
\bibliography{ref}

\end{document}